\documentclass{article}

\usepackage[utf8]{inputenc}
\usepackage{tismir}
\usepackage{amsmath}
\usepackage{hyperref}
\usepackage{url}
\usepackage{graphicx}
\usepackage{booktabs}

\usepackage{dsfont}
\usepackage{bookmark}
\usepackage{amssymb}
\usepackage[mathscr]{eucal}

\usepackage{subcaption}
\usepackage{tikz}
\usepackage{csquotes}
\usepackage{multirow}
\usepackage{wrapfig}
\usepackage[ruled,vlined]{algorithm2e}

\usepackage{todonotes}
\usepackage{stmaryrd}

\newcommand{\drawmatr}[6]{
    \draw[black,fill=#6] (#1,#2,#3) -- ++(#4,0,0) -- ++(0,-#5,0) -- ++(-#4,0,0) -- cycle;
}

\DeclareMathOperator*{\argmax}{arg\,max}

\newcommand{\writemetric}[2]{
$\genfrac{}{}{0pt}{}{#1}{#2}$
}

\newcommand{\ie}{\textit{i.e.}~}
\newcommand{\eg}{\textit{e.g.}~}

\DeclareMathOperator*{\Pz}{P_{0\text{bar}}}
\DeclareMathOperator*{\Rz}{R_{0\text{bar}}}
\DeclareMathOperator*{\Fz}{F_{0\text{bar}}}
\DeclareMathOperator*{\Po}{P_{1\text{bar}}}
\DeclareMathOperator*{\Ro}{R_{1\text{bar}}}
\DeclareMathOperator*{\Fo}{F_{1\text{bar}}}

\DeclareMathOperator*{\Pzf}{P_{0.5s}}
\DeclareMathOperator*{\Rzf}{R_{0.5s}}
\DeclareMathOperator*{\Fzf}{F_{0.5s}}
\DeclareMathOperator*{\Pth}{P_{3s}}
\DeclareMathOperator*{\Rth}{R_{3s}}
\DeclareMathOperator*{\Fth}{F_{3s}}

\newcommand{\TimeFrames}{N}

\newcommand{\BoundSet}{Z}
\newcommand{\aBound}{\zeta}
\newcommand{\AdmissibleSegmentations}{\Theta}
\newcommand{\aSeg}{S}

\newcommand{\SegScoreNotation}{u}
\newcommand{\SetOfScores}{\mathscr{U}^*}
\newcommand{\SetOfAntecedents}{A^*}

\definecolor{fondtitre}{RGB}{85,85,85}
\definecolor{fonddeboite}{RGB}{232,232,232}
\definecolor{shadecolor}{RGB}{232,232,232}

\usepackage{ntheorem}

\title{Barwise Music Structure Analysis with the Correlation Block-Matching Segmentation Algorithm}
\author{%
Axel Marmoret\thanks{Univ. Rennes 1, Inria, CNRS, IRISA, France.}\, \thanks{IMT Atlantique, Lab-STICC, Brest, France.},%
~Jérémy E. Cohen\thanks{CREATIS, Univ Lyon, CNRS, France.},%
~and Frédéric Bimbot\protect\footnotemark[1]}

\date{}

\authorref{Marmoret,~A., Cohen,~J.E., and Bimbot,~F.}
\journalyear{2023}
\journalvolume{6}
\journalissue{1}
\journalpages{ 167--185}
\doi{10.5334/tismir.167}

\authorshort{Marmoret,~A. et al.} 
\titleshort{Barwise Music Structure Analysis with the CBM Segmentation Algorithm}

\begin{document}


\twocolumn[{%
\maketitleblock
\begin{abstract}
Music Structure Analysis (MSA) is a Music Information Retrieval task consisting of representing a song in a simplified, organized manner by breaking it down into sections typically corresponding to ``chorus'', ``verse'', ``solo'', etc. In this work, we extend a MSA algorithm called the Correlation Block-Matching (CBM) algorithm introduced in~\citep{marmoret2020uncovering, marmoret2022barwise}. The CBM algorithm is a dynamic programming algorithm that segments self-similarity matrices, which are a standard description used in MSA and in numerous other applications. In this work, self-similarity matrices are computed from the feature representation of an audio signal and time is sampled at the barscale. This study examines three different standard similarity functions for the computation of self-similarity matrices. Results show that, in optimal conditions, the proposed algorithm achieves a level of performance which is competitive with supervised state-of-the-art methods while only requiring knowledge on bar positions.~In addition, the algorithm is made open-source and is highly customizable.
\end{abstract}
}]
\saythanks{}



\section{Introduction}
Citing \cite{paulus2010state}, ``[...] it is the structure, or the relationships between the sound events that create musical meaning''. In that sense, researchers in MIR developed the Music Structure Analysis (MSA) task, which focuses on the retrieval of the \textit{structure} in a song. Music structure is ill-defined, but is generally viewed as a hierarchical description, from the level of notes to the level of the song itself~\citep{mcfee2017evaluating, nieto2020segmentationreview}. A tentative definition is that structure is \textit{a simplified representation of the organization of the song}.

In that sense, motifs which arise from the organization of notes are a first level of structure.~These motifs create patterns, progressions and phrases. In general, the highest level of structure defines musical sections, corresponding to ``chorus'', ``verse'' and ``solo'', which is a macroscopic description of music~\citep{sargent2016estimating}.~Some work focus on estimating structure in its hierarchical nature, \eg\citep{mcfee2014analyzing, mcfee2014learning, de2020unveiling, salamon2021deep}, but this work focuses on a ``flat'' level of segmentation, \ie a macroscopic level, corresponding to musical sections. Facing the high diversity of music, and the many ways structure can be designed, we restrict this work to the study of Western modern (and in particular Western Popular) music. In particular, this work relies on both the RWC Pop~\citep{rwc} and the SALAMI~\citep{salami} datasets, which are open-source and standard datasets in MSA.

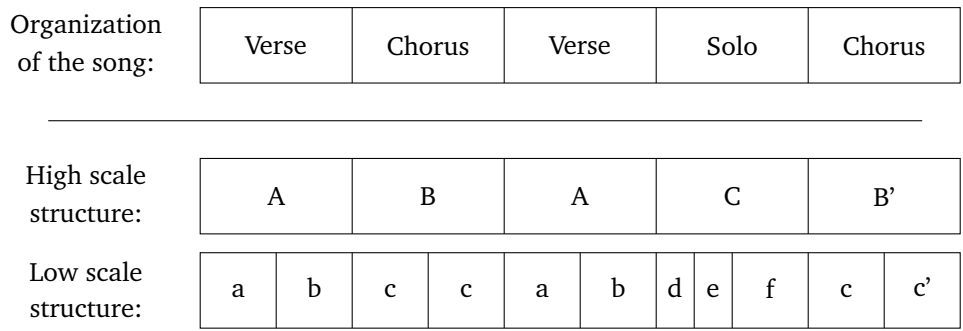
\begin{figure*}[htb!]
  \centering
  \begin{tikzpicture}
  \node at (-1.5,-0.25) {Organization};
  \node at (-1.5,-0.75) {of the song:};
  \drawmatr{0}{0}{0}{10}{1}{white};
  \draw[black] (2,0) -- ++(0,-1);
  \node at (1,-0.5) {Verse};
  \draw[black] (4,0) -- ++(0,-1);
  \node at (3,-0.5) {Chorus};
  \draw[black] (6,0) -- ++(0,-1);
  \node at (5,-0.5) {Verse};
  \draw[black] (8,0) -- ++(0,-1);
  \node at (7,-0.5) {Solo};
  \node at (9,-0.5) {Chorus};

  \draw[black] (-2,-1.5) -- ++(11.5,0);

  \node at (-1.5,-2.25) {High scale};
  \node at (-1.5,-2.75) {structure:};
  \drawmatr{0}{-2}{0}{10}{1}{white};
  \draw[black] (2,-2) -- ++(0,-1);
  \node at (1,-2.5) {A};
  \draw[black] (4,-2) -- ++(0,-1);
  \node at (3,-2.5) {B};
  \draw[black] (6,-2) -- ++(0,-1);
  \node at (5,-2.5) {A};
  \draw[black] (8,-2) -- ++(0,-1);
  \node at (7,-2.5) {C};
  \node at (9,-2.5) {B'};

  \node at (-1.5,-3.5) {Low scale};
  \node at (-1.5,-4) {structure:};
  \drawmatr{0}{-3.25}{0}{10}{1}{white};
  \draw[black] (1,-3.25) -- ++(0,-1);
  \node at (0.5,-3.755) {a};
  \draw[black] (2,-3.25) -- ++(0,-1);
  \node at (1.5,-3.71) {b};
  \draw[black] (3,-3.25) -- ++(0,-1);
  \node at (2.5,-3.75) {c};
  \draw[black] (4,-3.25) -- ++(0,-1);
  \node at (3.5,-3.75) {c};
  \draw[black] (5,-3.25) -- ++(0,-1);
  \node at (4.5,-3.755) {a};
  \draw[black] (6,-3.25) -- ++(0,-1);
  \node at (5.5,-3.71) {b};
  \draw[black] (6.5,-3.25) -- ++(0,-1);
  \node at (6.25,-3.715) {d};
  \draw[black] (7,-3.25) -- ++(0,-1);
  \node at (6.75,-3.757) {e};
  \draw[black] (8,-3.25) -- ++(0,-1);
  \node at (7.5,-3.72) {f};
  \draw[black] (9,-3.25) -- ++(0,-1);
  \node at (8.5,-3.75) {c};
  \node at (9.5,-3.70) {c'};
  \end{tikzpicture}
\caption{A schematic example of musical structure}
\label{fig:musical_structure}
\end{figure*}

MSA is subdivided into two subtasks, not necessarily mutually exclusive: the \textbf{boundary retrieval} task and the \textbf{segment labelling} task. The boundary retrieval task consists in estimating the boundaries between different sections, hence partitioning music in several non-overlapping segments, covering the entire song. The segment labelling task consists in grouping similar segments with a same label, typically letters such as 'A', 'B', 'C', etc. In this article, only the boundary retrieval task is considered. A schematic example of musical structure is presented in Figure~\ref{fig:musical_structure}.

\subsection{Related work}
\label{sec:sota}
Algorithms aimed at solving MSA are designed according to one or several criteria among the following: \textbf{homogeneity}, \textbf{novelty}, \textbf{repetition} and \textbf{regularity}~\citep{nieto2020segmentationreview}.~The homogeneity criterion assumes that a section consists of similar musical elements (notes, chords, tonality, timbre, ...). Novelty is the counterpart of homogeneity: this criterion considers that boundaries are located primarily between consecutive musical elements that are highly dissimilar.~A high novelty is salient between two distinct homogeneous zones (``break'' of homogeneity), and conversely, homogeneity is evaluated within successive dissimilar portions in the song. The third criterion, repetition, relies on a global approach of the song. The rationale is that a motif (\eg a melodic line) may be time-varying (and thus heterogeneous), but can define a segment if it is repeated across the song (for example, a chorus). The repetition criterion may also be used to partition long segments into smaller repeated segments. Finally, the regularity criterion assumes that, within a song (and even within a same musical genre), segments should be of comparable size.

Many MSA algorithms make use of matrices representing the similarity and dissimilarity in music, sometimes referred to as ``self-distance matrices''~\citep{paulus2010state}, ``self-similarity matrices''~\citep{nieto2020segmentationreview}, ``recurrence matrix'' or ``pair-wise frame similarities''~\citep{mcfee2014analyzing}. These representations differ in their details, but share the same conceptual idea of computing some form of similarity (or, conversely, some form of distance) between the different frames in music, and representing it in a square matrix (its size being the number of frames). In this work, we will use the term of ``self-similarity matrix''. 

As a particular example, the novelty kernel~\citep{foote2000automatic}, which may be one of the earliest work on audio MSA, estimates boundaries as points of high dissimilarity between the recent past and the near future, by applying a square kernel matrix on the diagonal of the self-similarity. 
This kernel works ideally when the recent past and the near future are homogenous in their respective neighborhoods, but very dissimilar with each other. 
In practice, this kernel is convolved with the self-similarity matrix of the song, centered on the diagonal, which gives a ``novelty'' value for each temporal sample of the song, finally post-processed into boundaries with a thresholding operation. 

While this technique is rather simple, it is still used as a standard segmentation tool in recent work, \eg~\citep{mccallum2019unsupervised, wang2021supervised} focusing on improving the boundary retrieval performance by enhancing the self-similarity matrix. In particular, both of these works belong to the domain of representation learning~\citep{bengio2013representation}, consisting of designing machine learning algorithms to learn relevant representations instead of focusing on solving a particular task.~In that context, using prior knowledge, both~\cite{mccallum2019unsupervised} and~\cite{wang2021supervised} design neural networks architectures and optimization schemes with the objective to obtain enhanced (nonlinear) similarity functions, more prone to highlight the structure in the self-similarity matrices.

Notably,~\cite{mccallum2019unsupervised} develops an unsupervised learning scheme where the prior knowledge enforced in the representation is based on the proximity of samples: the closer the frames in the song, the more probable they belong to the same segment. In the same spirit, ~\cite{wang2021supervised} develop a supervised learning scheme: the neural network learns representations where segments annotated with the same label are close, and segments annotated differently are far apart.~The rationale for both methods is to learn a similarity function which is not only representing the feature-wise correspondence of two music frames, but can also discover frequent patterns in the learning samples.

\cite{mcfee2014analyzing} propose an algorithm based on spectral clustering, aiming at interpreting the repetitive patterns in a song as principally connected vertices in a graph. The structure is then obtained by studying the eigenvectors of the Laplacian of this graph, forming cluster classes for segmentation. This technique is amongst the best-performing unsupervised techniques nowadays, and was improved by recent work by~\cite{salamon2021deep}, which replaces or enhances the acoustic features on which is applied spectral clustering with nonlinear embeddings, learned by means of a neural network.

\cite{serra2014unsupervised} develop ``Structural Features'', which, by design, encode both repetitive and homogeneous parts.~The rationale of these features is to compute the similarity between bags of instances, composed of several consecutive frames.~In that sense, the similarity encodes the repetition of any sequence, which can be stationary (homogeneity) or varying (repetition). Boundaries are obtained as points of high novelty between consecutive structural features.

Finally,~\cite{grill2015cnn} develop a Convolutional Neural Network (CNN) which outputs estimated boundaries. This CNN is one of the few techniques which does not compute a self-similarity to later post-process it into boundaries, but it still uses self-similarities as input. The network is supervised on two-level annotations, on the SALAMI dataset~\citep{salami}, and, according to the authors, using these two levels of annotations is beneficial to the performance. 

While many algorithms are devoted to the task of boundary retrieval (see for instance literature reviews from \cite{paulus2010state} and from~\cite{nieto2020segmentationreview}), research is still conducted towards more effective estimation algorithms. As presented above, in the past decade, research has mainly shifted from unsupervised to supervised algorithms, \ie from low-informed estimation algorithms, generally designed with strong hypotheses, to algorithms which take advantage from (generally huge) annotated databases to learn mappings between the musical features and annotated structural elements. While this shift has resulted in more effective algorithms, it has the disadvantages of requiring large training datasets, and reproducing potential bias in relation to the annotations, known to be prone to high subjectivity and ambiguity~\citep{nieto2020segmentationreview}. 

\subsection{Contributions}
In order to improve unsupervised algorithms, we propose in this article a novel approach based on the Correlation ``Block-Matching'' (CBM) algorithm. This algorithm was briefly introduced in previous work~\citep{marmoret2020uncovering, marmoret2022barwise} and is worth a more detailed presentation, which is one of the objectives of this work. Firstly, in line with the findings in~\citep{marmoret2020uncovering, marmoret2022barwise}, we conjecture that the barscale is the most appropriate temporal scale from which to infer structure in Western modern music, and we present a framework which inherently processes music at this temporal scale. To the best of our knowledge, only a few works used such an hypothesis (\eg\citep{wang2021supervised} and our previous works~\citep{marmoret2020uncovering, marmoret2022barwise}).
This hypothesis is supported by experiments which compare segmentation performance when aligning State-of-the-Art algorithms and the CBM algorithm on either the beat or the barscale. We show a consistent advantage for the barscale alignement approach.

The CBM algorithm estimates the musical structure based on the principles introduced in the work of~\cite{jensen2006multiple}, later extended by~\cite{sargent2016estimating}. In a nutshell, the CBM algorithm is based on the definition of a score function (further denoted as $\SegScoreNotation$) applied to segments, with the overall segmentation of the song resulting in the maximum total score of the set of segments. This defines an optimization problem, which can be solved by dynamic programming. 

The novelty of the CBM algorithm lies in its ability to extend previous work by incorporating new hypotheses regarding the design of the score function $\SegScoreNotation$. As a consequence, the algorithm is highly customizable and can be tailored to specific hypotheses and applications, which is a potential area for future research. We also present a study of different similarity functions to account for the similarity between musical features, and notably the Radial Basis function, which, to the best of our knowledge, was never previously used for MSA. Finally, we present experimental results which appear competitive with the most effective algorithm known to date~\citep{grill2015cnn}.

The CBM algorithm is unsupervised in the sense that the segment boundaries are estimated as solutions of an optimization problem which does not depend explicitly on annotated examples. Nonetheless, in order to accurately tune internal hyperparameters (and not empirically), the following experiments are carried out by separating data between a ``train'' and a ``test'' dataset. In addition, we acknowledge that we use a learning-based toolbox for the bar estimation, but this toolbox is independent form our work. In that sense, although we label this algorithm as ``unsupervised'', it could also arguably be qualified as ``weakly-'' or ``semi-'' supervised.

This article is organized as follows: Section~\ref{sec:barwise_music_analysis} presents in more details the hypotheses and framework to process music in a barwise setting, Section~\ref{sec:cbm_description} presents the CBM algorithm and Section~\ref{sec:experiments} presents an evaluation of the CBM algorithm on the boundary retrieval task, along with a comparison with State-of-the-Art algorithms.

\section{Barwise Music Analysis}
\label{sec:barwise_music_analysis}
In most work in MSA~\citep{nieto2020segmentationreview}, the signal of a song is represented as time-sampled features, related to some extent to the frequency content of the song's signal. In previous work on MSA, features have been either computed with a fixed hop length, typically between 0.1s and 1s according to~\cite{paulus2010state}, or (in more recent work), aligned on beats~\citep{mccallum2019unsupervised, wang2021supervised, salamon2021deep}. Beat alignment is musically-relevant because it aligns the features and the estimations with respect to a time segmentation consistent with music performance. In this work, we hypothesize that the barscale is more relevant than the beat scale to study MSA in Western modern music.

Bars seem well suited to express patterns and sections in Western modern music.~Indeed, in Western musical notations, musical notes lengths are expressed relatively to beats, and beats are combined to form bars. Bars finally segment the musical scores (with vertical lines), and similarities occur generally across different bars (which is particularly visible by the use of repeat bars, or symbols as ``Dal Segno'', ``Da Capo'', etc). In addition, the intuition that musical sections are synchronized on downbeats is experimentally confirmed by works such as~\citep{mauch2009using, fuentes2019music}, where the use of structural information improves the estimation of downbeats. Experiments supporting this hypothesis are presented in Section~\ref{sec:experiments_barwise_sota}.

The direct drawback of barwise alignment is the need for a powerful tool to estimate bars. In this work, we use the \textit{madmom} toolbox~\citep{madmom}, which uses a neural network to perform bar estimation~\citep{bock2016joint}. In the 2016 MIREX contest\endnote{\href{https://www.music-ir.org/mirex/wiki/2016:Audio_Downbeat_Estimation_Results}{www.music-ir.org/mirex/wiki/2016:Audio\\\_Downbeat\_Estimation\_Results}}, which was the last edition of the contest comparing downbeat estimation algorithms, this neural network obtained the best performance, and can hence be considered as one of the State-of-the-Art algorithms for the task. Even if some algorithms obtained better performance since (\eg\citep{bock2020deconstruct, oyama2021phase, hung2022modeling}), we consider that the \textit{madmom} toolbox achieves a satisfactory level of performance for our intended application.

\subsection{Barwise TF Matrix}
In this work, we represent music as barwise spectrograms, and more particularly as a \textbf{Barwise TF matrix}, following~\citep{marmoret2022barwise}. 
The Barwise TF matrix consists of a matrix of size $B \times TF$, $B$ being the number of bars in the song (\ie a dimension accounting for the barscale), and $TF$ the vectorization of both time (at barscale) and feature dimensions (representing the frequency to some extent) into a unique Time-Frequency dimension. The number of time frames per bar is fixed to $T = 96$, as in~\citep{marmoret2022barwise}. Following the work of~\cite{grill2015cnn}, the signal is represented in log mel features, \ie the logarithm of mel coefficients, expressed with $F = 80$ mel coefficients, but any other feature representation could be used instead. The rationale for using log mel spectrograms is that they lead to high segmentation performance~\citep{grill2015cnn, nieto2020segmentationreview} while constituting a compact spectral representation, suited for music analysis.

\subsection{Barwise Self-Similarity Matrix}
\label{sec:self_similarity}

\begin{figure}[tb!]
  \centering
  \includegraphics[width=0.8\columnwidth]{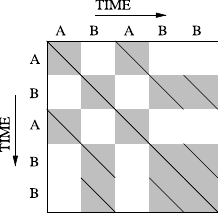}
  \caption{An idealized self-similarity matrix, extracted from~\citep{paulus2010state}.}
  \label{fig:self_similarity_example_sota}
\end{figure}

As stated in Section~\ref{sec:sota}, a common representation in MSA is the self-similarity matrix, representing the similarities at the scale of the song. An idealized self-similarity matrix, extracted from~\citep{paulus2010state}, is presented in Figure~\ref{fig:self_similarity_example_sota}. Similar passages are identified by two typical shapes: \textbf{blocks} and \textbf{stripes}. A block is a square (or a rectangle) in the self-similarity, representing a zone of high inner-similarity, \ie several consecutive frames which are highly similar, hence corresponding to the homogeneity criterion. A stripe is a line parallel to the main diagonal representing a repetition of the content, \ie a pattern of several frames repeated in the same order, hence corresponding to the repetition criterion. As a general trend, the segmentation algorithms using self-similarity matrices are designed so as to retrieve segments based on blocks and stripes.

Given a Barwise TF matrix $X \in \mathbb{R}^{B\times TF}$, the self-similarity matrix of $X$ is defined as $A(X) \in \mathbb{R}^{B\times B}$ where each coefficient $(i, j)$ represents the similarity between vectors $X_i, X_j \in \mathbb{R}^{TF}$. Self-similarity matrices are computed from the Barwise TF representation of the song, therefore, each coefficient in the self-similarity represents the feature-wise similarity for a pair of bars.

The similarity between two vectors is subject to a similarity function (the dot product for instance), and, as a consequence, different self-similarity matrices can be constructed. 
The main diagonal in a self-similarity matrix represents the self-similarity of each vector, and is in general (and in this work in particular) normalized to one. This work studies three different similarity functions, namely the Cosine, Autocorrelation and RBF similarity functions. The latter two represent novel contributions compared to our previous work~\citep{marmoret2022barwise}.

\begin{figure*}[htb!]
\centering
  \includegraphics[width=2\columnwidth]{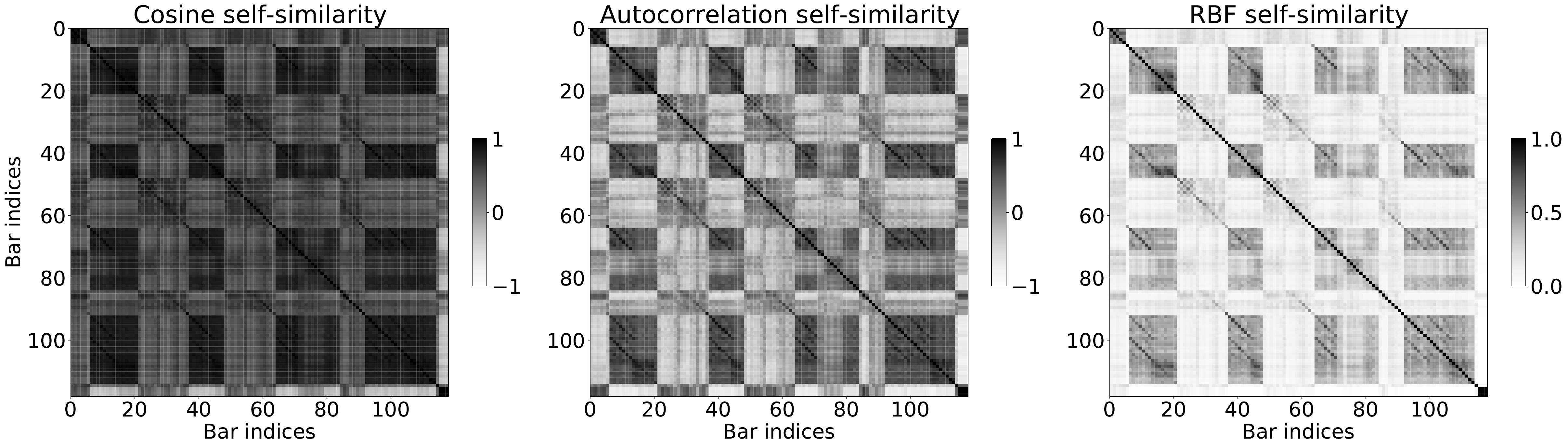}
 \caption{Cosine, Autocorrelation and RBF self-similarities on the song \textit{POP01} of RWC Pop.}
\label{fig:different_autosimilarities}
\end{figure*}

\subsubsection{Cosine Self-Similarity Matrix}
The Cosine similarity function computes the normalized dot products between two vectors, and leads to the Cosine self-similarity matrix, denoted as $A_{cos}(X)$. Practically, denoting as $\tilde{X}$ the row-wise $l_2$-normalized version of $X$ (\ie the matrix $X$ where each row has been divided by its $l_2$-norm), the Cosine self-similarity matrix is defined as $A_{cos}(X) = \tilde{X}\tilde{X}^{\intercal}$, or, elementwise, for $1 \leq i, j \leq B$:
\begin{equation}
A_{cos}(X)_{ij} = \frac{\left\langle X_i, X_j\right\rangle}{\|X_i\|_2 \|X_j\|_2} = \sum_{k = 1}^{TF} \tilde{X}_{ik} \tilde{X}_{jk}.
\end{equation}


\subsubsection{Autocorrelation Self-Similarity Matrix}
The Autocorrelation similarity function is defined for 2 bars $X_i$ and $X_j$ as $corr(X_i, X_j)~=~(X_i - \bar{x})(X_j - \bar{x})^{\intercal}$, denoting as $\bar{x} \in \mathbb{R}^{TF}$ the mean of all bars in the song.

The barwise Autocorrelation similarity function yields the Autocorrelation self-similarity matrix $A_{corr}(X)$ as:
\begin{equation}
  A_{corr}(X)_{ij} = \frac{\left\langle X_i - \bar{x}, X_j - \bar{x} \right\rangle}{\|X_i - \bar{x}\|_2 \|X_j - \bar{x}\|_2}.
\end{equation}
In other words, the Autocorrelation matrix is exactly the Cosine self-similarity matrix of the centered matrix $X~-~\mathds{1}_B^{\intercal}~\bar{x}$, \ie $A_{corr}(X)~=~A_{cos}\left(X - \mathds{1}_B^{\intercal} \bar{x}\right)$.

\subsubsection{RBF Self-Similarity Matrix}
Kernel functions are symmetric positive definite or semi-definite functions.~In machine learning, kernel functions are generally used to represent data in a high-dimensional space (sometimes infinite), enabling a nonlinear processing of data with linear methods (\eg nonlinear classification with Support Vector Machines, SVM).

The Radial Basis Function (RBF) kernel is a kernel function defined as $\text{RBF}(X_i, X_j)~=~\exp(-\gamma\| X_i - X_j\|^2_2)$, $\gamma$ being a user-defined parameter. The RBF can be used as a similarity function between two bars $X_i$ and $X_j$, hence defining the RBF self-similarity matrix $A_{RBF}(X)$ as:
\begin{equation}
A_{RBF}(X)_{ij} = \text{RBF}(\tilde{X_i}, \tilde{X_j}) = \exp\left(-\gamma\left\|\frac{X_i}{\|X_i\|_2} - \frac{X_j}{\|X_j\|_2}\right\|^2_2\right).
\end{equation}
Bars are normalized by their $l_2$ norm in the computation of $A_{RBF}$, in order to limit the impact of variations of power between bars. The self-similarity of a bar is equal to $e^0~=~1$.

Parameter $\gamma$ is set relatively to the standard deviation of the pairwise Euclidean distances of all bars in the original matrix (self-distances excluded), to adapt the shape of the exponential function to the relative distribution of distances in the song.
Hence, denoting as: $\sigma~=~\underset{\underset{i \neq j}{1 \leq i,j \leq B}}{std}\left(\left\|\frac{X_i}{\|X_i\|_2} - \frac{X_j}{\|X_j\|_2}\right\|^2_2 \right)$, we set $\gamma~=~\frac{1}{2\sigma}$.

The RBF function may be useful for MSA as the self-similarity of dissimilar elements fades rapidly due to the properties of the exponential function. Hence, the RBF similarity function emphasizes on similar components (\ie on homogeneous zones).

The three self-similarities are presented in Figure~\ref{fig:different_autosimilarities}, on the Barwise TF of song \textit{POP01} from RWC Pop.

\subsection{Barwise MSA Experiments}
\label{sec:experiments_barwise_sota}
Section~\ref{sec:barwise_music_analysis} is based on the hypothesis that the barscale is more relevant than other time discretizations (in particular the beat scale) to study MSA in Western modern music. To support this hypothesis, we present hereafter three experiments studying the differences in performance between beat-aligned and barwise-aligned estimations.

\subsubsection{Aligning the annotations on the downbeats}
\begin{table*}[htb!]
  \centering
  \begin{tabular}{llllllll}
  \hline
  \multicolumn{2}{l} {Dataset} & $\Pzf$ & $\Rzf$ & $\Fzf$ & $\Pth$ & $\Rth$ & $\Fth$ \\ \hline
  SALAMI & Annotation 1  & 82.47\% & 82.14\% & 82.30\% & 99.94\% & 99.56\% & 99.74\% \\ 
              & Annotation 2  & 80.97\% & 80.92\% & 80.94\% & 99.92\% & 99.84\% & 99.88\% \\
  \multicolumn{2}{l} {RWC Pop} & 96.46\% & 96.21\% & 96.33\% & 100\% & 99.73\% & 99.86\% \\ \hline
  \end{tabular}
  \caption{Standard metrics (see Section~\ref{sec:metrics}) when aligning the reference annotations on the downbeats (compared to the original annotations).}
  \label{table:ref_aligned_downbeats}
\end{table*}

As a preliminary experiment to study the impact of barwise alignment on the segmentation quality, we evaluate the loss in performance when aligning annotations on downbeats for both the RWC Pop~\citep{rwc} and SALAMI~\citep{salami} datasets. In this experiment, each annotation is aligned with the closest estimated downbeat, and the barwise-aligned annotations are compared with the initial annotation using the standard metrics $\Pzf$, $\Rzf$, $\Fzf$ and $\Pth$, $\Rth$, $\Fth$ (detailed in Section~\ref{sec:metrics}). Results are presented in Table~\ref{table:ref_aligned_downbeats}. The RWC Pop annotations are barely impacted by the barwise-alignement, suggesting that annotations are precisely located on downbeats.~
A loss in performance with short tolerances is observed on the annotations of the SALAMI dataset, either suggesting imprecise bar estimations or boundaries not located on the downbeats. Still, the levels of performance exhibited in Table~\ref{table:ref_aligned_downbeats} largely outperform the nowadays State-of-the-Art ($\approx 80\%$ \textit{vs}~$\approx 54\%$ for the $\Fzf$ metric, respectively for the downbeat-aligned annotations in Table~\ref{table:ref_aligned_downbeats} and for~\citep{grill2015cnn}, whose results are presented in Figure~\ref{fig:comparison_SOTA_salami}). In that sense, the loss in performance induced by downbeat alignment may be compensated if estimations are indeed more precise due to this alignment.

\subsubsection{Downbeat-Alignment for Several State-of-the-Art Algorithms}
\label{sec:downbeat_alignment}
\begin{figure}[htb!]
\centering
\begin{subfigure}{\columnwidth}
  \includegraphics[width=\columnwidth]{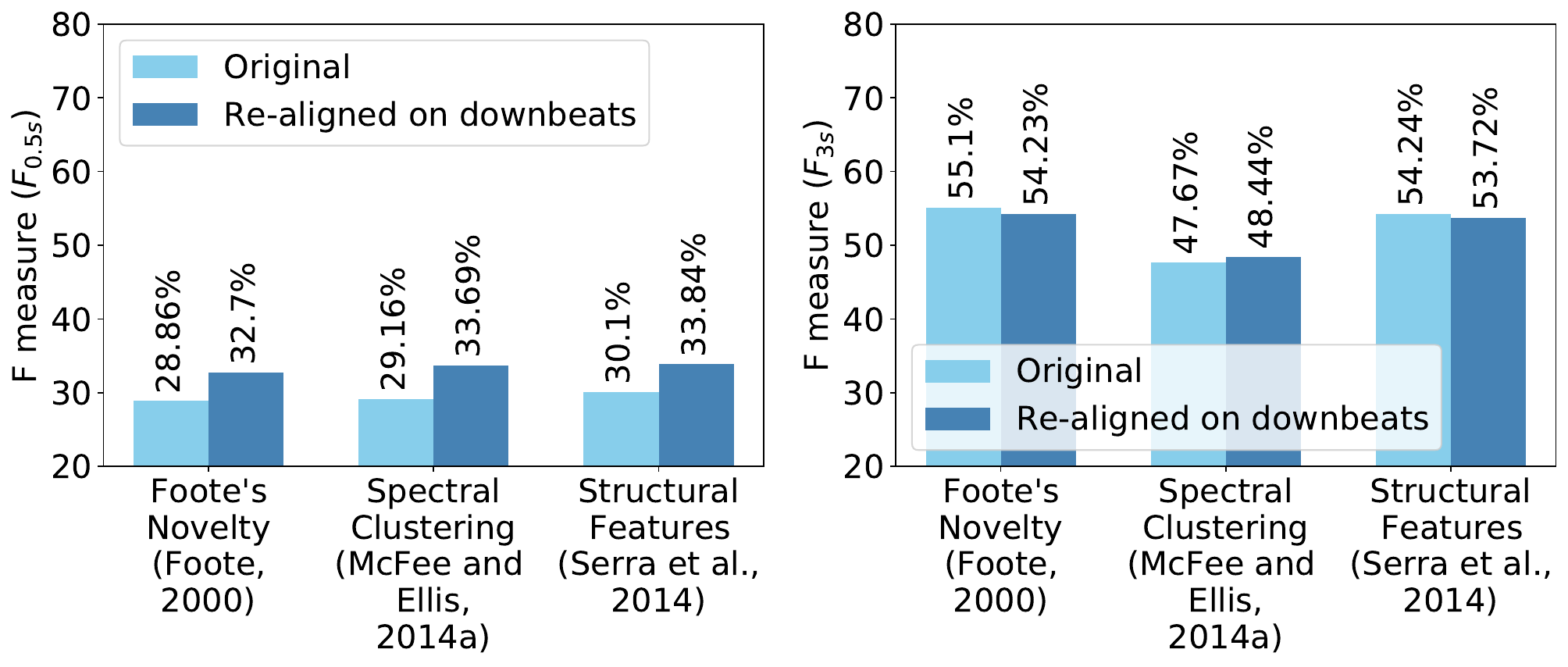}
  \caption{SALAMI-test.}
  \vspace{10pt}
  \label{fig:results_sota_salami_beat_bar}
  \end{subfigure}
  \quad
\begin{subfigure}{\columnwidth}
  \includegraphics[width=\columnwidth]{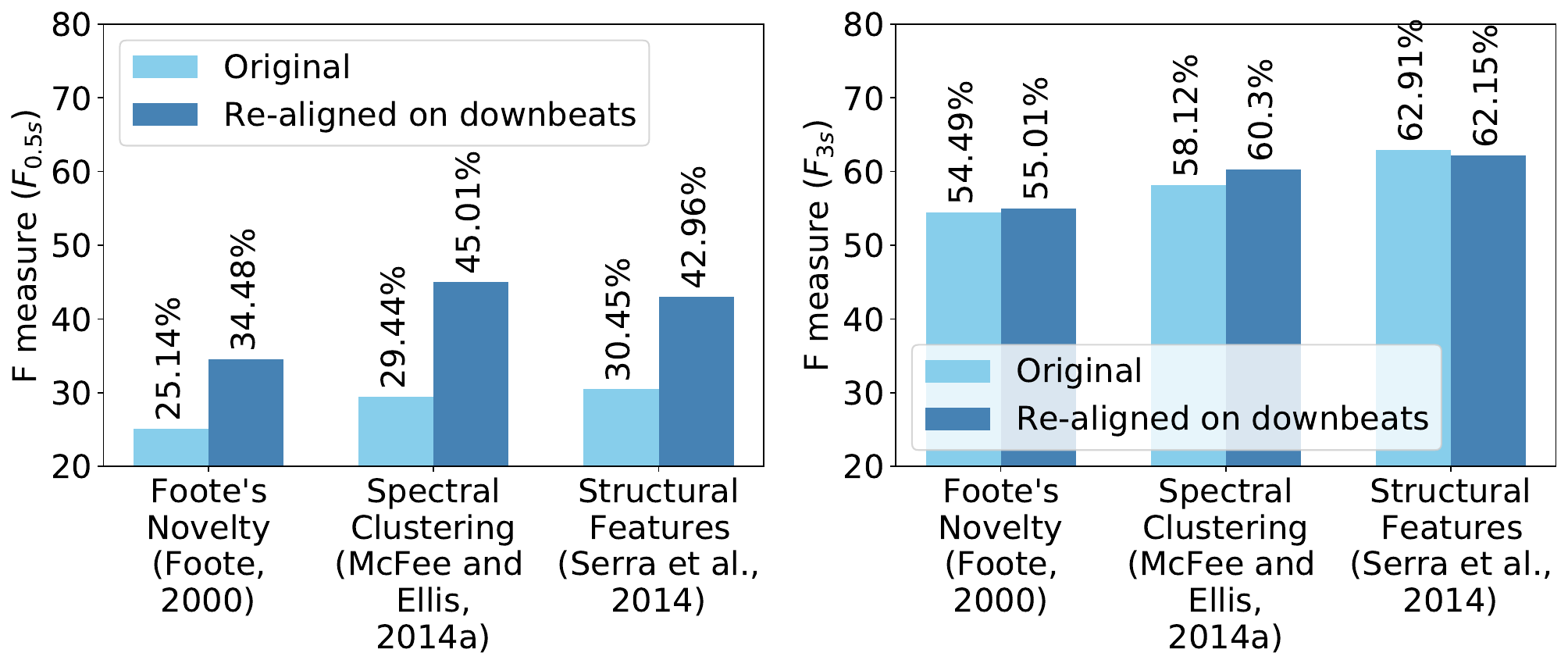}
  \caption{RWC Pop.}
  \label{fig:results_sota_rwc_beat_bar}
\end{subfigure}
\caption{Segmentation results of State-of-the-Art algorithms on the SALAMI-test and RWC Pop datasets, for beat-aligned (original) \textit{vs.} downbeat-aligned boundaries. The SALAMI-test dataset is defined in~\citep{ullrich2014boundary}, and introduced in Section~\ref{sec:train_test_datasets}.}
\label{fig:results_sota_beat_bar}
\end{figure}

A second experiment consists of post-processing the boundary estimations of three unsupervised State-of-the-Art algorithms~\citep{foote2000automatic, mcfee2014analyzing, serra2014unsupervised}, computed with the \textit{MSAF} toolbox~\citep{msaf}, by aligning each boundary with the closest estimated downbeat. As these algorithms originally use beat-aligned features (resulting in beat-aligned estimations), this experiment compares beat-aligned estimations with downbeat-aligned estimations. Segmentation scores are presented in Figure~\ref{fig:results_sota_beat_bar} for both SALAMI and RWC Pop datasets. 

Results show that aligning estimated boundaries on downbeats results in a strong increase in performance for $\Fzf$, and to comparable results for $\Fth$, on both datasets. Hence, aligned on downbeats, estimations are more accurate, but the $\Fth$ metric is not significantly impacted by this alignment. These results suggest that downbeat-alignment is beneficial on these datasets.

\subsubsection{Focusing on Foote's algorithm}
\label{sec:foote_details}
Finally, a third experiment compares the results obtained with different time discretizations for Foote's algorithm~\citep{foote2000automatic}, implemented in the \textit{MSAF} toolbox~\citep{msaf}.~In particular, this experiment compares the results when self-similarities are computed with beat-aligned and downbeat-aligned features.

As presented in Section~\ref{sec:sota}, Foote's algorithm estimates boundaries as points of high novelty. A novelty score is computed at each point of the self-similarity matrix by applying a kernel matrix, and this novelty score is finally post-processed into boundaries with a thresholding operation.~In the original implementation, the cosine self-similarity is computed on beat-synchronized features, \ie one feature per beat.~In this experiment, beats are estimated with the algorithm of~\cite{bock2019multi}, which is one of the State-of-the-Art algorithms in beat estimation, and is implemented in the \textit{madmom} toolbox.

As in Section~\ref{sec:downbeat_alignment}, the original results are compared with the ones obtained when aligning the estimations on downbeats. In addition, we compare the beat-synchronized results with two new feature processing: \textbf{bar-synchronized features}, \ie one feature per bar (instead of one per beat), and the \textbf{Barwise TF matrix}\endnote{In this experiment, the Barwise TF matrix is computed on the same features than in the implementation of the algorithm of~\cite{foote2000automatic} in \textit{MSAF} (\ie chroma features).}, introduced in Section~\ref{sec:self_similarity}. 

The kernel is of size 66 for the beat-synchronized features (as originally set in \textit{MSAF}), and is of size 16 for both the bar-synchronized features and the Barwise TF matrix.~Results are presented in Tables~\ref{table:foote_beat_bar_salami} and~\ref{table:foote_beat_bar_rwc}, respectively for the SALAMI-test and the RWC Pop dataset. In order to fairly compare the algorithms, we also fitted two hyperpameters of the original \textit{MSAF} implementation for the barscale, namely the size of median filtering applied to the input spectrogram and the standard deviation in the gaussian filter applied to the novelty curve. These parameters were fitted in a train/test fashion, as detailed in Section~\ref{sec:train_test_datasets}.


\begin{table*}[htb!]
  \centering
  
  \begin{tabular}{llllllll}
  \hline 
    \multicolumn{2}{l} {Time synchronization} & $\Pzf$ & $\Rzf$ & $\Fzf$ & $\Pth$ & $\Rth$ & $\Fth$ \\ \hline
  Beat-synchronized & Original & 26.98\% & 34.58\% & 29.21\% & 50.10\% & 63.30\% & 54.02\% \\
   		& Re-aligned on downbeats  & 31.05\% & 39.15\% & 33.33\% & 50.08\% & 62.95\% & 53.78\% \\
  \multicolumn{2}{l} {Bar-synchronized} & 37.68\% & 36.36\% & 35.97\% & 58.06\% & 56.11\% & 55.57\% \\
  \multicolumn{2}{l} {Barwise TF Matrix} & \textbf{39.22\%} &\textbf{42.66\%} & \textbf{39.67\%} & \textbf{59.60\%} & \textbf{64.82\%} & \textbf{60.36\%} \\ \hline
   \end{tabular}
  
  \caption{Different time synchronizations for the~\cite{foote2000automatic} algorithm on the SALAMI-test dataset. The SALAMI-test dataset is defined in~\citep{ullrich2014boundary}, and introduced in Section~\ref{sec:train_test_datasets}.}
  \label{table:foote_beat_bar_salami}
\end{table*}

\begin{table*}[htb!]
  \centering
  
  \begin{tabular}{llllllll}
  \hline
    \multicolumn{2}{l} {Time synchronization} & $\Pzf$ & $\Rzf$ & $\Fzf$ & $\Pth$ & $\Rth$ & $\Fth$ \\ \hline
  Beat-synchronized & Original & 31.86\% & 24.38\% & 27.29\% & 67.21\% & 51.92\% & 57.95\% \\
   				    & Re-aligned on downbeats  & 42.30\% & 32.82\% & 36.52\% & 66.67\% & 51.44\% & 57.44\% \\
  \multicolumn{2}{l} {Bar-synchronized} & 43.53\% & 26.32\% & 32.46\% & 69.25\% & 42.22\% & 51.97\% \\
  \multicolumn{2}{l} {Barwise TF Matrix} & \textbf{53.09\%} & \textbf{37.19\%} & \textbf{43.30\%} & \textbf{79.35\%} & \textbf{56.03\%} & \textbf{65.04\%} \\ \hline  
  \end{tabular}
  
  \caption{Different time synchronizations for the~\cite{foote2000automatic} algorithm on the RWC Pop dataset.}
  \label{table:foote_beat_bar_rwc}
\end{table*}

Both Tables~\ref{table:foote_beat_bar_salami} and~\ref{table:foote_beat_bar_rwc} conclude in the same direction: the best performance for the $\Fzf$ and $\Fth$ metrics are obtained with the Barwise TF matrix. Similarly than in Section~\ref{sec:downbeat_alignment}, aligning beat-aligned estimations on downbeats in post-processing increases the performance for the $\Fzf$ metric, indicating more precise estimations. It is worthwile noting that, using bar-synchronized instead of beat-synchronized features increases the performance on the SALAMI dataset, while it decreases it on the RWC Pop dataset. 

However, Barwise TF matrix representation appears to be  beneficial for MSA on both datasets. In future experiments, we denote as ``Foote-TF'' the condition where Foote's algorithm is applied to the Barwise-TF matrix, whose results are shown in Tables~\ref{table:foote_beat_bar_salami} and~\ref{table:foote_beat_bar_rwc}.


\section{Correlation ``Block-Matching'' Algorithm}\label{sec:cbm_description}
With a self-similarity matrix as input, the Correlation ``Block-Matching'' segmentation algorithm (CBM) estimates boundaries by means of dynamic programming. This algorithm is detailed in this section, along with a study of important parameter settings which were not discussed in the previous work~\citep{marmoret2020uncovering,marmoret2022barwise}, such as the block weighting kernels and the penalty functions. The CBM algorithm estimates boundaries based on the homogeneity/novelty and regularity criteria. 
The principles of dynamic programming are presented in a first part, and the definition of a score function $\SegScoreNotation$ applied on segments is detailed in a second part.

\subsection{Dynamic Programming for Boundary Retrieval}
\label{sec:dynaprog_definition}

\subsubsection{Boundary Retrieval Problem}
Given a music piece (song) sampled in time as $\TimeFrames$ time steps, the subtask of boundary retrieval can be defined as finding a segmentation (set of boundaries) $\BoundSet$ representing the start of each segment, \ie $\BoundSet~=~\left\{\aBound_i \in \llbracket 1, \TimeFrames \rrbracket, \, i \in \llbracket 1, E \rrbracket\right\}$, $E$ representing the number of boundaries estimated in this song. The set of admissible segmentations is denoted as $\AdmissibleSegmentations$, \ie $\BoundSet \in \AdmissibleSegmentations$.~Each segment $\aSeg_i$ is composed of the time steps between two consecutive boundaries, \ie $\aSeg_i~=~\left\{l \in \llbracket 1, \TimeFrames \rrbracket \mid \aBound_i \leq l < \aBound_{i+1}\right\}$. The second bound is exclusive as it represents the start of the next segment $\aSeg_{i+1}$. 
By definition, $E$ boundaries define $E-1$ segments. Boundary $\aBound_{i}$ is called the \textbf{antecedent} of boundary $\aBound_{i+1}$.

\subsubsection{Barwise Boundary Retrieval Problem}
In the proposed barwise paradigm, the song is discretized into $B$ bars using $N = B+1$ bar boundaries. Hence, the first boundary is the start of the song, \ie $\aBound_1 = 1$, the last boundary is the end of the last bar in the song\endnote{As the song contains $B$ bars, $B+1$ represents the end of the last bar, \ie the last downbeat of the song.}, \ie $\aBound_E = B+1$, and each boundary is located on a bar, \ie $\forall i, \, \aBound_i \in \llbracket 1, B+1 \rrbracket$. Each segment $\aSeg_i$ is composed of the bar indices between two consecutive boundaries.

As a consequence, there exists\endnote{As each set of boundaries must contain the first and last downbeats of the song, at most $2^{B-1}$ sets of boundaries can be obtained.} $\binom{B-1}{E}$ different sets of boundaries composed of exactly $E$ boundaries, and, more generally, at most $\sum\limits_{k=0}^{B-1} \binom{B-1}{k} = 2^{B-1}$ segmentations for each song. Hence, the segmentation problem admits a finite number of solutions, which can theoretically be solved in a combinatorial way. In practice though, evaluating all possible segmentations leads to an algorithm of exponential complexity $\mathcal{O}\left(2^B\right)$, considered intractable in practice.

\subsubsection{Dynamic Programming}
The boundary retrieval problem can be approached as an optimization problem~\citep{jensen2006multiple, sargent2016estimating}. In particular, by associating a score $\SegScoreNotation(\aSeg)$ to each potential segment $\aSeg$, the optimal segmentation $\BoundSet^*$ is the segmentation maximizing\endnote{In details, both~\cite{jensen2006multiple} and~\cite{sargent2016estimating} introduced the optimal segmentation as the minimum of a cost function, when it is rather defined here as a maximum. It actually depends on the way of conceiving the score function $\SegScoreNotation$, and, in particular, by defining a cost function equal to the inverse of the score function $\SegScoreNotation$, both problems are equivalent.} the sum of all its segment scores:
\begin{equation}
\begin{aligned}
  \BoundSet^* &= \underset{\BoundSet \in \AdmissibleSegmentations}{\argmax} \quad \sum\limits_{i=1}^{E-1} \SegScoreNotation\left(\llbracket \aBound_i, \aBound_{i+1} - 1\rrbracket\right)
  \\
  &=\underset{\BoundSet \in \AdmissibleSegmentations}{\argmax} \quad \SegScoreNotation(\BoundSet)
\end{aligned}
  \label{eq:optimal_segmentation}
\end{equation}

by extending notation $\SegScoreNotation$ for a set of segments.

The problem can be solved using a dynamic programming algorithm~\citep[Chap. 15]{bellman1952theory, cormen2009introduction}, which principle is to solve a combinatorial optimization problem by dividing it into several independent subproblems. The independent subproblems are formulated in a recursive manner, and their solutions can be stitched together to form a solution to the original problem. 
Notice that in the current formulation of the segmentation problem, defined in Equation~\ref{eq:optimal_segmentation}, each potential segment is evaluated independently, via its score, and is never compared with the others. In other terms, repetitions of the same section are not considered, while they could inform on the overall structure, typically considering the repetition criterion. Thus, the segmentation problem defined in Equation~\ref{eq:optimal_segmentation} is a relaxation of the general segmentation problem.~This relaxation is considered because it allows to use principles of dynamic programming, by evaluating the score of all segments as independent subproblems. In particular, this relaxed problem is said to exhibit ``optimal substructure''~\citep{cormen2009introduction}.

\subsubsection{Longest-Path on a Directed Acyclic Graph}
Following the formulation of~\cite{jensen2006multiple}, the segmentation problem can be reframed into the problem of finding the longest path on a Directed Acyclic Graph (DAG).~
The rationale of the solution algorithm is that the optimal segmentation up to any given bar $b_k$ 
can be found exactly by recursively evaluating the optimal segmentations up to each antecedent of $b_k$, \ie (without any constraint) all bars $b_l < b_k$, and the score of the segments $\llbracket b_l, b_k - 1 \rrbracket$. Formally, denoting as $\BoundSet^*_{[1:b_k]}$ the optimal segmentation up to bar $b_k$, the CBM algorithm consists of:
\begin{enumerate}
  \item Lookup for $\left\{\SegScoreNotation\left(\BoundSet^*_{[1:b_l]}\right) , \, \forall b_l < b_k\right\}$, \ie the 
  optimal segmentation up to each antecedent, which is stored in an array when first computed,
  \item Computing $\left\{\SegScoreNotation\left(\llbracket b_l, b_k - 1 \rrbracket\right) , \, \forall b_l < b_k\right\}$, \ie the 
  segmentation score between 
  bars $b_l$ and $b_k$,
  \item Finding the best antecedent of $b_k$, denoted as $\aBound_{b_k-1}^*$, with the following equation:
  \begin{equation}
    \aBound_{b_k-1}^* = \underset{b_l}{\argmax}\left(\SegScoreNotation\left(\BoundSet^*_{[1:b_l]}\right) + \SegScoreNotation\left(\llbracket b_l, b_k - 1 \rrbracket\right)\right).
    \label{eq:recurrence_property_dynaprog}
  \end{equation}
\end{enumerate}

\begin{figure*}[htb!]
\centering
\begin{tikzpicture}[scale=1]
\node[draw,circle] (b1) at (-6,0) {$b_1$};
\node[draw,circle] (b2) at (-1,3) {$b_2$};
\node[draw,circle] (b3) at (1,-1) {$b_3$};
\node[draw,circle] (b4) at (6,1.5) {$b_4$};

\draw[->,>=latex, line width=0.5mm] (b1) -- (b2) node[near end](b1b2){};
\draw[->,>=latex, line width=0.5mm] (b1) -- (b3) node[midway](b1b3){};
\draw[->,>=latex] (b1) -- (b4) node[near end](b1b4){};
\draw[->,>=latex] (b2) -- (b3) node[near start](b2b3){};
\draw[->,>=latex] (b2) -- (b4) node[near start](b2b4){};
\draw[->,>=latex, line width=0.5mm] (b3) -- (b4) node[midway](b3b4){};

\node[above=6mm of b1.west] {$\SegScoreNotation\left(\BoundSet^*_{[b_1:b_1]}\right) = 0$};

\node[left=1.5mm of b1b2] {2};
\node[right=5mm of b1b3.north] {6};
\node[above=0.1mm of b1b4] {8};

\node[left=5mm of b2] (setb2) {$\BoundSet^*_{[b_1:b_2]} = \{b_1,b_2\}$};
\node[above=1mm of setb2.north]  {$\SegScoreNotation\left(\BoundSet^*_{[b_1:b_2]}\right) = \SegScoreNotation\left(\llbracket b_1, b_2 -1 \rrbracket\right) = 2$};

\node[left=0.1mm of b2b3] {1};
\node[right=4mm of b2b4] {1};

\node[below=1mm of b3] (scoreSetb3) {$\SegScoreNotation\left(\BoundSet^*_{[b_1:b_3]}\right) = \text{max}\left(\SegScoreNotation\left(\BoundSet^*_{[b_1:b_2]}\right) + \SegScoreNotation\left(\llbracket b_2, b_3 - 1\rrbracket\right), \SegScoreNotation\left(\llbracket b_1, b_3 - 1\rrbracket\right)\right) = 6$};
\node[right=1mm of b3] {$\BoundSet^*_{[b_1:b_3]} = \{b_1,b_3\}$};

\node[left=1.5mm of b3b4] {4};

\node[above=13mm of b4.north](scoreSetb4){$\SegScoreNotation\left(\BoundSet^*_{[b_1:b_4]}\right) = \SegScoreNotation\left(\BoundSet^*_{[b_1:b_3]}\right)  + \SegScoreNotation\left(\llbracket b_3, b_4 - 1 \rrbracket\right) = 10$};

\node[right=3mm of b4.west](BoundSetb4Line2) {\qquad$= \{b_1,b_3,b_4\}$};
\node[above=2mm of BoundSetb4Line2](BoundSetb4Line1) {$\BoundSet^*_{[b_1:b_4]} = \BoundSet^*_{[b_1:b_3]}\cup \{b_4\}$};

\end{tikzpicture}
 \caption{Computing the optimal segmentation with 4 bars.}
 \label{fig:recurrence_property_dynaprog}
\end{figure*}
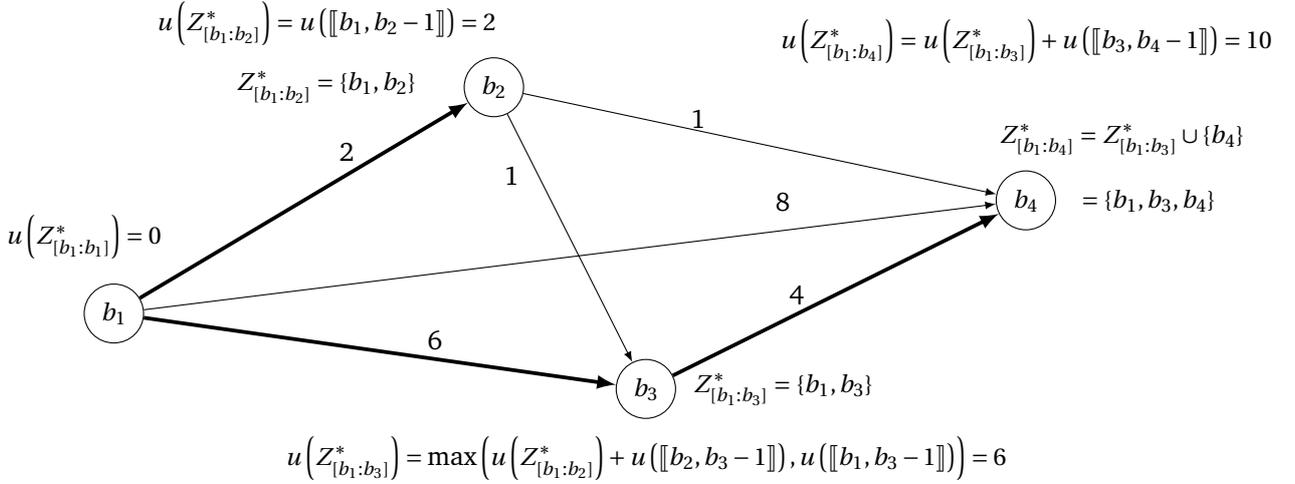

Finally, at the last iteration, the algorithm computes the best antecedent for $B+1$, \ie the last downbeat of the song. Then, recursively, the algorithm is able to backtrack the best antecedent of this antecedent, and so on and so forth back to the first bar of the song, thus providing the optimal segmentation. A graph visualization for a 4-bar example is presented in Figure~\ref{fig:recurrence_property_dynaprog}. A pseudo-code for the CBM algorithm, assuming that the score function $\SegScoreNotation$ is given, is detailed in appendix (Algorithm~\ref{alg:dyna_conv_algo}).

In the end, for any bar $b_k$, the optimal segmentation up to $b_k$ can be computed in $\mathcal{O}\left(b_k - 1\right)$ operations, \ie parsing each antecedent only once. Hence, the solution algorithm boils down to $\mathcal{O}\left(\frac{B(B+1)}{2}\right)$ evaluations, which corresponds to a polynomial complexity. 
In practice, we even limit the size of admissible segments to be at most 32 bars (set empirically), which further reduces the complexity. 

\subsection{Score Function}
Finally, the segmentation problem boils down to the definition of the score function $\SegScoreNotation\left(\llbracket \aBound_i, \aBound_{i+1} - 1\rrbracket\right)$ for a segment. In the CBM algorithm and following~\citep{sargent2016estimating}, the score of each segment is defined as a mixed score function, presented in Equation~\ref{eq:mixed_score_function} as the balanced sum of two terms:
\begin{equation}
  \SegScoreNotation\left(\llbracket \aBound_i, \aBound_{i+1} - 1\rrbracket\right) = \SegScoreNotation^{K}\left(\llbracket \aBound_i, \aBound_{i+1} - 1\rrbracket\right) - \lambda p\left(\aBound_{i+1} - \aBound_i\right).
  \label{eq:mixed_score_function}
\end{equation}

The first term, $\SegScoreNotation^{K}\left(\llbracket \aBound_i, \aBound_{i+1} - 1 \rrbracket\right)$, is based on the homogeneity criterion, and is presented in Section~\ref{sec:convolutional_kernels}; the second one, $p\left(\aBound_{i+1} - \aBound_i\right)$, is based on the regularity criterion, and is presented in Section~\ref{sec:penalty_function}. Parameter $\lambda$ is a balancing parameter.

\subsubsection{Block Weighting Kernels}
\label{sec:convolutional_kernels}
The first term $\SegScoreNotation^{K}$ of the score function in Equation~\ref{eq:mixed_score_function} is obtained from the self-similarity values within a segment.~Practically, given a self-similarity matrix $A(X)$, the score $\SegScoreNotation^{K}(\aSeg_i)$ of segment $\aSeg_i~=~\llbracket~\aBound_{i},~\aBound_{i+1}~-~1\rrbracket$ (of size $n~=~\aBound_{i+1}~-~\aBound_{i}$) is computed by evaluating the self-similarity values restricted to $\aSeg_i$, \ie $A\left(X_{\aSeg_i}\right)~=~A\left(X_{[\aBound_{i}:\aBound_{i+1} - 1]}\right)$. It can be understood as cropping the self-similarity $A(X)$ on this particular segment, around the diagonal.

The CBM algorithm aims at favoring the homogeneity of estimated segments, \ie favoring sections composed of similar elements. Thus, the score function $\SegScoreNotation^{K}$ is defined so as to measure the inner similarity of a segment. In practice, this is obtained through weighting local self-similarity values, by using a (fixed) weighting kernel matrix $K$, such as: 

\begin{equation}
\begin{array}{ccccc}
  \SegScoreNotation^{K} & : & \mathbb{R}^{n \times n} & \to & \mathbb{R} \\
  & & A\left(X_{\aSeg_i}\right) & \mapsto & \frac{1}{n} \sum\limits_{k = 1}^{n} \sum\limits_{l = 1}^{n} A\left(X_{\aSeg_i}\right)_{kl} K_{kl}. \\
\end{array}
\end{equation}

The kernel is called a ``weighting kernel''. A first observation is that the weighting kernel needs to adapt to the size of the segment. A very simple kernel is a kernel matrix full of ones, \ie $K = \mathds{1}_{n\times n}$, resulting in a score function equal to the sum of every element in the self-similarity, normalized by the size of the segment. The normalization by the size of the segment is meant to turn the squared dependence of the size of the segment in the number of self-similarity values ($n^2$ values in the self-similarity) into a linear dependence. A linear dependence is desired as it ensures a length-$n$ segment contributes similarly to the sum of segment scores as $n$ segments of length 1.

The design of the weighting kernel defines how to transform bar similarities into segment homogeneity, which is of particular importance for segmentation. The remainder of this section presents two types of kernels, namely the ``full'' kernel and the ``band'' kernel. We consider that the main diagonal in the self-similarity is not informative regarding the overall similarity in the segment, as its values are normalized to one. Hence, for every weighting kernel $K$ used in the CBM algorithm, $K_{ii} = 0, \, \forall i$.

\paragraph{Full Kernel}
The first kernel is called the 
``full'' kernel, because it corresponds to a kernel full of 1 (except on the diagonal where it is equal to 0). The full kernel captures the average value of similarities in this segment, excluding the self-similarity values. Practically, denoting as $K^{f}$ the full kernel: \begin{equation}
  K^{f}_{ij} = \begin{cases} 1 & \text{if } i \neq j \\ 0 & \text{if } i = j \end{cases}
\end{equation}

\begin{figure}[htb!]
  \begin{center}
    \begin{tikzpicture}[scale=2.05]
      \drawmatr{0}{0}{0}{1}{1}{black}
      \drawmatr{0}{-0.1}{0}{0.1}{0.4}{black}
      \drawmatr{0}{0}{0}{0.1}{0.1}{white}
      \drawmatr{0.1}{-0.1}{0}{0.1}{0.1}{white}
      \drawmatr{0.2}{-0.2}{0}{0.1}{0.1}{white}
      \drawmatr{0.3}{-0.3}{0}{0.1}{0.1}{white}
      \drawmatr{0.4}{-0.4}{0}{0.1}{0.1}{white}
      \drawmatr{0.5}{-0.5}{0}{0.1}{0.1}{white}
      \drawmatr{0.6}{-0.6}{0}{0.1}{0.1}{white}
      \drawmatr{0.7}{-0.7}{0}{0.1}{0.1}{white}
      \drawmatr{0.8}{-0.8}{0}{0.1}{0.1}{white}
      \drawmatr{0.9}{-0.9}{0}{0.1}{0.1}{white}

      \drawmatr{1.1}{-0.4}{0}{0.1}{0.1}{black}
      \node at (1.3,-0.45) {1};
      \drawmatr{1.1}{-0.6}{0}{0.1}{0.1}{white}
      \node at (1.3,-0.65) {0};
\end{tikzpicture}
\end{center}
  \caption{Full kernel of size 10}
  \label{fig:full_kernel}
\end{figure}
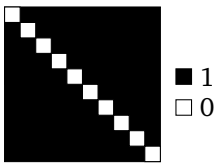

Hence, the score function associated with the full kernel is equal to:
\begin{equation}
  \SegScoreNotation^{K^{f}}(\aSeg_i) = \frac{1}{n} \sum\limits_{k = 1}^{n} \sum\limits_{l = 1}^{n} A\left(X_{\aSeg_i}\right)_{kl} K^{f}_{kl} = \frac{1}{n} \sum\limits_{k = 1}^{n} \sum\limits_{l = 1, l \neq k}^{n} A\left(X_{\aSeg_i}\right)_{kl}
\end{equation}
A full kernel of size 10 is presented in Figure~\ref{fig:full_kernel}.

\paragraph{Band Kernels}
A second class of kernels, called ``band'' kernels, are considered in order to emphasize short-term similarity. Indeed, in band kernels, the weighting score is computed on the pairwise similarities of a few bars in the segment only, depending on their temporal proximity: only close bars are considered. In practice, this can be obtained by defining a kernel with entries equal to 0, except on some upper- and sub-diagonals. The number of upper- and sub-diagonals is a parameter, corresponding to the maximal number of bars considered to evaluate the similarity, \ie an upper bound on $|b_i-b_j|$ for a pair of bars $(b_i, b_j)$.~

Hence, a band kernel is defined according to its number of bands, denoted as $v$, defining the $v$-band kernel $K^{vb}$ such that:
\begin{equation}
\begin{array}{crl}
K^{vb}_{ij} = \begin{cases} 1 & \text{if } 1 \leq |i - j| \leq v, \\ 0 & \text{otherwise } (i = j \text{ or } |i - j| > v). \end{cases}
\end{array}
\end{equation}

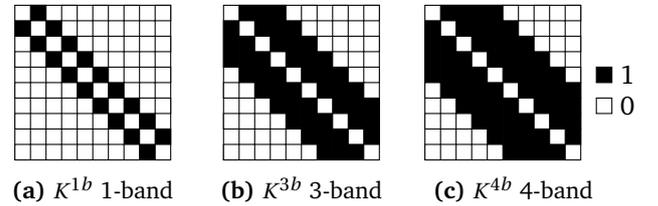
\begin{figure}[htb!]
  \centering
  \begin{subfigure}{0.27\columnwidth}
  \centering

    \begin{tikzpicture}[scale=2.05]
    \drawmatr{0}{0}{0}{1}{1}{white}

    \drawmatr{0}{-0.1}{0}{0.1}{0.1}{black}
    \drawmatr{0.1}{-0.2}{0}{0.1}{0.1}{black}
    \drawmatr{0.2}{-0.3}{0}{0.1}{0.1}{black}
    \drawmatr{0.3}{-0.4}{0}{0.1}{0.1}{black}
    \drawmatr{0.4}{-0.5}{0}{0.1}{0.1}{black}
    \drawmatr{0.5}{-0.6}{0}{0.1}{0.1}{black}
    \drawmatr{0.6}{-0.7}{0}{0.1}{0.1}{black}
    \drawmatr{0.7}{-0.8}{0}{0.1}{0.1}{black}
    \drawmatr{0.8}{-0.9}{0}{0.1}{0.1}{black}

    \drawmatr{0.1}{0}{0}{0.1}{0.1}{black}
    \drawmatr{0.2}{-0.1}{0}{0.1}{0.1}{black}
    \drawmatr{0.3}{-0.2}{0}{0.1}{0.1}{black}
    \drawmatr{0.4}{-0.3}{0}{0.1}{0.1}{black}
    \drawmatr{0.5}{-0.4}{0}{0.1}{0.1}{black}
    \drawmatr{0.6}{-0.5}{0}{0.1}{0.1}{black}
    \drawmatr{0.7}{-0.6}{0}{0.1}{0.1}{black}
    \drawmatr{0.8}{-0.7}{0}{0.1}{0.1}{black}
    \drawmatr{0.9}{-0.8}{0}{0.1}{0.1}{black}

    \drawmatr{0}{-0.3}{0}{0.1}{0.1}{white}
    \drawmatr{0.1}{-0.4}{0}{0.1}{0.1}{white}
    \drawmatr{0.2}{-0.5}{0}{0.1}{0.1}{white}
    \drawmatr{0.3}{-0.6}{0}{0.1}{0.1}{white}
    \drawmatr{0.4}{-0.7}{0}{0.1}{0.1}{white}
    \drawmatr{0.5}{-0.8}{0}{0.1}{0.1}{white}
    \drawmatr{0.6}{-0.9}{0}{0.1}{0.1}{white}

    \drawmatr{0.3}{0}{0}{0.1}{0.1}{white}
    \drawmatr{0.4}{-0.1}{0}{0.1}{0.1}{white}
    \drawmatr{0.5}{-0.2}{0}{0.1}{0.1}{white}
    \drawmatr{0.6}{-0.3}{0}{0.1}{0.1}{white}
    \drawmatr{0.7}{-0.4}{0}{0.1}{0.1}{white}
    \drawmatr{0.8}{-0.5}{0}{0.1}{0.1}{white}
    \drawmatr{0.9}{-0.6}{0}{0.1}{0.1}{white}

    \drawmatr{0}{-0.5}{0}{0.1}{0.1}{white}
    \drawmatr{0.1}{-0.6}{0}{0.1}{0.1}{white}
    \drawmatr{0.2}{-0.7}{0}{0.1}{0.1}{white}
    \drawmatr{0.3}{-0.8}{0}{0.1}{0.1}{white}
    \drawmatr{0.4}{-0.9}{0}{0.1}{0.1}{white}

    \drawmatr{0.5}{0}{0}{0.1}{0.1}{white}
    \drawmatr{0.6}{-0.1}{0}{0.1}{0.1}{white}
    \drawmatr{0.7}{-0.2}{0}{0.1}{0.1}{white}
    \drawmatr{0.8}{-0.3}{0}{0.1}{0.1}{white}
    \drawmatr{0.9}{-0.4}{0}{0.1}{0.1}{white}

    \drawmatr{0}{-0.7}{0}{0.1}{0.1}{white}
    \drawmatr{0.1}{-0.8}{0}{0.1}{0.1}{white}
    \drawmatr{0.2}{-0.9}{0}{0.1}{0.1}{white}

    \drawmatr{0.7}{0}{0}{0.1}{0.1}{white}
    \drawmatr{0.8}{-0.1}{0}{0.1}{0.1}{white}
    \drawmatr{0.9}{-0.2}{0}{0.1}{0.1}{white}

    \drawmatr{0}{-0.9}{0}{0.1}{0.1}{white}

    \drawmatr{0.9}{0}{0}{0.1}{0.1}{white}


  \end{tikzpicture}
    \caption{$K^{1b}$ 1-band}
  \end{subfigure}
  \quad
  \begin{subfigure}{0.28\columnwidth}
    \centering

    \begin{tikzpicture}[scale=2.05]
    \drawmatr{0}{0}{0}{1}{1}{white}

    \drawmatr{0}{-0.1}{0}{0.1}{0.3}{black}
    \drawmatr{0.1}{-0.2}{0}{0.1}{0.3}{black}
    \drawmatr{0.2}{-0.3}{0}{0.1}{0.3}{black}
    \drawmatr{0.3}{-0.4}{0}{0.1}{0.3}{black}
    \drawmatr{0.4}{-0.5}{0}{0.1}{0.3}{black}
    \drawmatr{0.5}{-0.6}{0}{0.1}{0.3}{black}
    \drawmatr{0.6}{-0.7}{0}{0.1}{0.3}{black}
    \drawmatr{0.7}{-0.8}{0}{0.1}{0.2}{black}
    \drawmatr{0.8}{-0.9}{0}{0.1}{0.1}{black}

    \drawmatr{0.1}{0}{0}{0.3}{0.1}{black}
    \drawmatr{0.2}{-0.1}{0}{0.3}{0.1}{black}
    \drawmatr{0.3}{-0.2}{0}{0.3}{0.1}{black}
    \drawmatr{0.4}{-0.3}{0}{0.3}{0.1}{black}
    \drawmatr{0.5}{-0.4}{0}{0.3}{0.1}{black}
    \drawmatr{0.6}{-0.5}{0}{0.3}{0.1}{black}
    \drawmatr{0.7}{-0.6}{0}{0.3}{0.1}{black}
    \drawmatr{0.8}{-0.7}{0}{0.2}{0.1}{black}
    \drawmatr{0.9}{-0.8}{0}{0.1}{0.1}{black}

    \drawmatr{0}{-0.5}{0}{0.1}{0.1}{white}
    \drawmatr{0.1}{-0.6}{0}{0.1}{0.1}{white}
    \drawmatr{0.2}{-0.7}{0}{0.1}{0.1}{white}
    \drawmatr{0.3}{-0.8}{0}{0.1}{0.1}{white}
    \drawmatr{0.4}{-0.9}{0}{0.1}{0.1}{white}

    \drawmatr{0.5}{0}{0}{0.1}{0.1}{white}
    \drawmatr{0.6}{-0.1}{0}{0.1}{0.1}{white}
    \drawmatr{0.7}{-0.2}{0}{0.1}{0.1}{white}
    \drawmatr{0.8}{-0.3}{0}{0.1}{0.1}{white}
    \drawmatr{0.9}{-0.4}{0}{0.1}{0.1}{white}

    \drawmatr{0}{-0.7}{0}{0.1}{0.1}{white}
    \drawmatr{0.1}{-0.8}{0}{0.1}{0.1}{white}
    \drawmatr{0.2}{-0.9}{0}{0.1}{0.1}{white}

    \drawmatr{0.7}{0}{0}{0.1}{0.1}{white}
    \drawmatr{0.8}{-0.1}{0}{0.1}{0.1}{white}
    \drawmatr{0.9}{-0.2}{0}{0.1}{0.1}{white}

    \drawmatr{0}{-0.9}{0}{0.1}{0.1}{white}

    \drawmatr{0.9}{0}{0}{0.1}{0.1}{white}

  \end{tikzpicture}
    \caption{$K^{3b}$ 3-band}
    \end{subfigure}
    \quad
    \begin{subfigure}{0.28\columnwidth}
      \centering

      \begin{tikzpicture}[scale=2.05]
      \drawmatr{0}{0}{0}{1}{1}{white}

      \drawmatr{0}{-0.1}{0}{0.1}{0.4}{black}
      \drawmatr{0.1}{-0.2}{0}{0.1}{0.4}{black}
      \drawmatr{0.2}{-0.3}{0}{0.1}{0.4}{black}
      \drawmatr{0.3}{-0.4}{0}{0.1}{0.4}{black}
      \drawmatr{0.4}{-0.5}{0}{0.1}{0.4}{black}
      \drawmatr{0.5}{-0.6}{0}{0.1}{0.4}{black}
      \drawmatr{0.6}{-0.7}{0}{0.1}{0.3}{black}
      \drawmatr{0.7}{-0.8}{0}{0.1}{0.2}{black}
      \drawmatr{0.8}{-0.9}{0}{0.1}{0.1}{black}

      \drawmatr{0.1}{0}{0}{0.4}{0.1}{black}
      \drawmatr{0.2}{-0.1}{0}{0.4}{0.1}{black}
      \drawmatr{0.3}{-0.2}{0}{0.4}{0.1}{black}
      \drawmatr{0.4}{-0.3}{0}{0.4}{0.1}{black}
      \drawmatr{0.5}{-0.4}{0}{0.4}{0.1}{black}
      \drawmatr{0.6}{-0.5}{0}{0.4}{0.1}{black}
      \drawmatr{0.7}{-0.6}{0}{0.3}{0.1}{black}
      \drawmatr{0.8}{-0.7}{0}{0.2}{0.1}{black}
      \drawmatr{0.9}{-0.8}{0}{0.1}{0.1}{black}

      \drawmatr{0}{-0.5}{0}{0.1}{0.1}{white}
      \drawmatr{0.1}{-0.6}{0}{0.1}{0.1}{white}
      \drawmatr{0.2}{-0.7}{0}{0.1}{0.1}{white}
      \drawmatr{0.3}{-0.8}{0}{0.1}{0.1}{white}
      \drawmatr{0.4}{-0.9}{0}{0.1}{0.1}{white}

      \drawmatr{0.5}{0}{0}{0.1}{0.1}{white}
      \drawmatr{0.6}{-0.1}{0}{0.1}{0.1}{white}
      \drawmatr{0.7}{-0.2}{0}{0.1}{0.1}{white}
      \drawmatr{0.8}{-0.3}{0}{0.1}{0.1}{white}
      \drawmatr{0.9}{-0.4}{0}{0.1}{0.1}{white}

      \drawmatr{0}{-0.7}{0}{0.1}{0.1}{white}
      \drawmatr{0.1}{-0.8}{0}{0.1}{0.1}{white}
      \drawmatr{0.2}{-0.9}{0}{0.1}{0.1}{white}

      \drawmatr{0.7}{0}{0}{0.1}{0.1}{white}
      \drawmatr{0.8}{-0.1}{0}{0.1}{0.1}{white}
      \drawmatr{0.9}{-0.2}{0}{0.1}{0.1}{white}

      \drawmatr{0}{-0.9}{0}{0.1}{0.1}{white}

      \drawmatr{0.9}{0}{0}{0.1}{0.1}{white}

      \drawmatr{1.1}{-0.4}{0}{0.1}{0.1}{black}
      \node at (1.3,-0.45) {1};
      \drawmatr{1.1}{-0.6}{0}{0.1}{0.1}{white}
      \node at (1.3,-0.65) {0};
    \end{tikzpicture}
      \caption{$K^{4b}$ 4-band}
      \end{subfigure}
          \vspace{10pt}
  \caption{Band kernels, of size 10}
  \label{fig:bands_kernel}
\end{figure}

Three band kernels, of size 10, are represented in Figure~\ref{fig:bands_kernel}. Section~\ref{sec:experiments} presents experiments which compare quantitatively the impact of the number of bands on the segmentation performance.

\subsubsection{Penalty Functions}
\label{sec:penalty_function}
\cite{sargent2016estimating} extended the score function of~\cite{jensen2006multiple} to take into account both the homogeneity and the regularity criteria, resulting in Equation~\ref{eq:mixed_score_function}.
In practice, this is obtained through defining a regularity penalty function $p(n)$, corresponding to the second term in Equation~\ref{eq:mixed_score_function}, and penalizing segments according to their size $n$, to favor particular sizes. 

\begin{figure}[htb!]
\centering
\begin{subfigure}{0.98\columnwidth}
\centering
  \includegraphics[width=0.9\columnwidth]{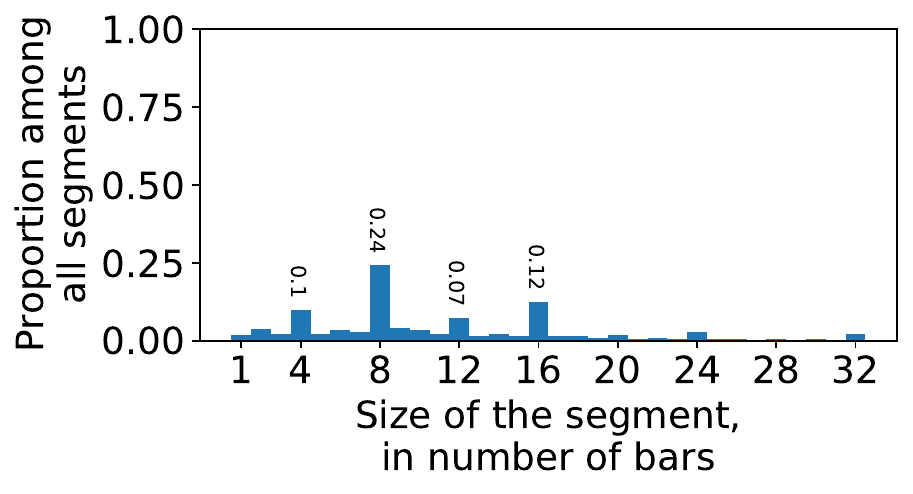}
    \vspace{-2pt}
  \caption{SALAMI-test.}
  \vspace{7pt}
\end{subfigure}
\quad
\begin{subfigure}{0.98\columnwidth}
\centering
  \includegraphics[width=0.9\columnwidth]{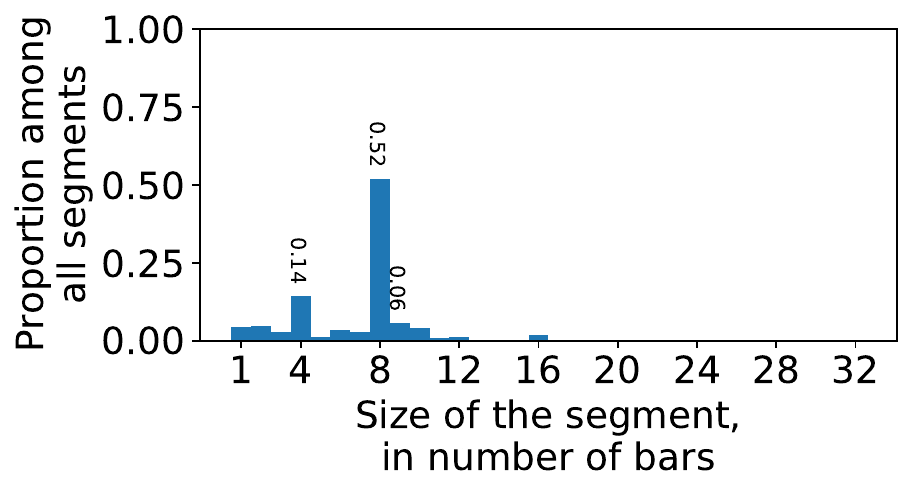}
  \caption{RWC Pop.}
    \vspace{4pt}
  \end{subfigure}
\caption{Distribution of segment sizes in terms of number of bars, in the annotations.}
\label{fig:distrib_barwise_segment_sizes_cbm_annotation}
\end{figure}

The penalty function is based on prior knowledge, and aims at enforcing particular sizes of segments, which are known to be typical in a number of music genres, notably Pop music. In particular, Figure~\ref{fig:distrib_barwise_segment_sizes_cbm_annotation} presents the distributions of the sizes of segments, in terms of number of bars, in the annotations of both RWC Pop and SALAMI datasets. It appears that some sizes of segments are much more frequent in the annotations. Hence, penalty functions $p$ can be derived from these distributions.

Two different penalty functions $p$ are studied in this section, namely the ``target-deviation'' and ``modulo'' functions. In what follows, $n$ denotes the size of the segment, \ie $n = \aBound_{i+1} - \aBound_i$.

\paragraph{Target-Deviation Functions}
The first set of penalty functions, called ``target-deviation'' and denoted as $p^{td}$, is defined by~\cite{sargent2016estimating}.~Target-deviation functions compute the difference between the size of the current estimated segment and a target size $\tau$, raised to the power of a parameter $\alpha$, \ie $p^{td}(n) = |n - \tau|^{\alpha}$ where parameter $\alpha$ takes typical values in $\{0.5, 1, 2\}$. The target size is set by~\cite{sargent2016estimating} to 32, to favor segments of size 32 beats, in line with their respective evaluations of most frequent segment sizes. In our barwise context\endnote{Note that 8 bars containing 4 beats each leads to 32 beats. 4 beats per bar is a frequent value for Western Popular music.}, $\tau = 8$, which is the most frequent segment size in both RWC Pop and SALAMI datasets.

This penalty function is adapted to enforce one size in particular, and tends to disadvantage all the others. Hence, this function is adapted to datasets where one size is predominant, which seems true for RWC Pop with MIREX 10 annotations (more than half of the segments in the annotation are of size 8 bars), but not so definite for the SALAMI dataset, where the segment sizes are more balanced between 4, 8, 12 and 16, as presented in Figure~\ref{fig:distrib_barwise_segment_sizes_cbm_annotation}. In particular, segments of size 16 are strongly penalized ($|8-16|^{\alpha}~=~8^{\alpha}$). 

\paragraph{Modulo function}
The second set of penalty functions, called ``modulo functions'', is designed to favor particular segment sizes, directly based on prior knowledge. In this study, we only present the ``modulo 8'' function $p^{m8}(n)$ based on both RWC Pop and SALAMI annotations. Indeed, in both datasets, most segments are of size 8, and the remaining segments are generally of size 4, 12 or 16. Finally, outside of these sizes, even segments are more frequent than segments of odd sizes. Hence, the modulo 8 function models this distribution, as:
\begin{equation}
 p^{m8}(n) = \begin{cases}
  0 & \text{if } n = 8 \\
  \frac{1}{4} & \text{else, if } n \equiv 0 \pmod 4 \\
  \frac{1}{2} & \text{else, if } n \equiv 0 \pmod 2 \\
   1 & \text{otherwise}
 \end{cases}
\label{eq:modulo_8}
\end{equation}
Penalty values for the different cases were set quite intuitively, and would benefit from further investigation.

In order to mitigate both the weighting score function $\SegScoreNotation^{K}$ and the penalty function $p$, we implemented an additional normalization step based on the weighted values obtained in each song, resulting in the score function defined in Equation~\ref{eq:detailed_score_func}.
\begin{equation}
  \SegScoreNotation\left(\llbracket \aBound_i, \aBound_{i+1} - 1 \rrbracket\right) = \SegScoreNotation^{K}\left(\llbracket \aBound_i, \aBound_{i+1} - 1\rrbracket\right) - \SegScoreNotation^{K8}_{\text{max}}\lambda p\left(\aBound_{i+1} - \aBound_i\right),
  \label{eq:detailed_score_func}
\end{equation}
In Equation~\ref{eq:detailed_score_func}, $\SegScoreNotation^{K8}_{\text{max}}$ is the maximal weighting value obtained by sliding a kernel of size 8 on this self-similarity, \ie the highest score among all possible segments of size 8. This size of 8 for the kernel is chosen as the most frequent segment size in terms of number of bars in both RWC Pop and SALAMI datasets, as presented in Figure~\ref{fig:distrib_barwise_segment_sizes_cbm_annotation}. Parameter $\lambda$ is a constant parameter, which is fitted as detailed in Section~\ref{sec:experiments}.

Finally, in the CBM algorithm, the score of each segment is defined as in Equation~\ref{eq:detailed_score_func}. The first term, $\SegScoreNotation^{K}\left(\llbracket \aBound_i, \aBound_{i+1} - 1\rrbracket\right)$, is a weighting score, measuring the self similarity of the segment. The second term, $p\left(\aBound_{i+1} - \aBound_i\right)$, penalizes or favors the segment depending on its size. Both these scores are subject to design choices, which are studied and compared in the subsequent section.

\section{Experiments}
\label{sec:experiments}
\subsection{Evaluation Metrics}
\label{sec:metrics}
The quality of the estimation obtained with the CBM algorithm is evaluated with the Hit-Rate metrics, comparing a set of estimated boundaries with a set of annotations by intersecting them with respect to a tolerance~$t$~\citep{ong2005semantic, turnbull2007supervised}.~In practice, given two sets of boundaries $\BoundSet^e$ and $\BoundSet^a$ (respectively the sets of estimated and annotated boundaries), an estimated boundary $\aBound^e_i \in \BoundSet^e$ is considered correct if it is close enough to an annotated boundary $\aBound^a_j \in \BoundSet^a$ (``close enough'' meaning that the gap is no larger than the tolerance $t$), \ie if $\exists\, \aBound^a_j \in \BoundSet^a \,/\, \left|\aBound^e_i - \aBound^a_j\right| \leq t$.
Each estimated boundary can be coupled with a maximum of one annotated boundary, and \textit{vice versa}.~
The set of correct boundaries subject to the tolerance $t$, denoted as $C_t$, contains at most as many elements as the annotations or the estimations, \ie $0 \leq |C_t| \leq \text{min}\left(|\BoundSet^e|, |\BoundSet^a|\right)$. In case of perfect concordance between $\BoundSet^e$ and $\BoundSet^a$, $C_t = \BoundSet^e = \BoundSet^a$. In practice, the concordance of $C_t$ with $\BoundSet^e$ and $\BoundSet^a$ is evaluated by the precision $P_t$, recall $R_t$ and F-measure $F_t$:
\begin{itemize}
  \item $P_t = \frac{|C_t|}{|\BoundSet^e|}$, \ie the proportion of accurately estimated boundaries among the total number of estimated boundaries. 
  \item $R_t = \frac{|C_t|}{|\BoundSet^a|}$, \ie the proportion of accurately estimated boundaries among the total number of annotated boundaries. 
  \item $F_t = \frac{2 P_t R_t}{P_t + R_t}$ is the harmonic mean of both aforementioned measures. The harmonic mean is less sensible to large values than the arithmetic (standard) mean, and is conversely more strongly penalized by low values. Hence, a high F-measure requires both a high recall and a high precision.
\end{itemize}
These metrics are computed using the \textit{mir\_eval} toolbox~\citep{mireval}.

\subsubsection{Tolerances in Absolute Time}
In the boundary retrieval subtask, conventions for the tolerance values are 0.5s~\citep{turnbull2007supervised} and 3s~\citep{ong2005semantic}. The 3 seconds tolerance, citing~\cite{ong2005semantic}, is justified as being equal to ``approximately 1 bar for a song of quadruple meter [NB: 4 beats per bar, \eg\writemetric{4}{4} metric] with 80 bpm in tempo'', while the 0.5 second tolerance is within the order of magnitude corresponding to the beat. In this work, we use both tolerance values to compare our algorithm with the standard algorithms, leading to 6 metrics $\Pzf$, $\Rzf$, $\Fzf$ and $\Pth$, $\Rth$, $\Fth$. In these metrics, estimations are compared with the original annotations.

\subsubsection{Barwise-Aligned Tolerances}
In this work, estimated boundaries are located on downbeat estimations, as explained and motivated in Section~\ref{sec:barwise_music_analysis}. In that sense, rather than evaluating the estimations in absolute time, 
we align each annotation with the closest estimated downbeat, leading to barwise-aligned annotations. This allows us to introduce additional metrics: $\Pz$, $\Rz$, $\Fz$ and $\Po$, $\Ro$, $\Fo$. The first three metrics (\eg $\Fz$) consider that the tolerance is set to 0 bar, \ie expecting estimations and annotations to fall precisely on the same downbeat, and the latter three metrics (\eg $\Fo$) set the tolerance to exactly one bar between estimations and annotations. In particular, these metrics will be used to compare different settings of our algorithm.

\subsection{Parametrization of the Algorithm}
\label{sec:parametrization}
The CBM algorithm is evaluated on the boundary retrieval task on the entire RWC Pop dataset~\citep{rwc}, and on the test subset of SALAMI~\citep{salami}, defined by~\cite{ullrich2014boundary}.~
The three similarity functions defined in Section~\ref{sec:self_similarity} are used to compute the self-similarity matrices, namely the Cosine, Autocorrelation and RBF. The CBM algorithm itself is subject to the choice of the kernel, and particularly to the number of bands when using a band kernel. In addition, the score function depends on the design of the penalty function. Rather than studying all of these parameters at the same time, experiments focus on each aspect independently. In particular, the experiments aim at answering the three following questions:

\begin{itemize}
\item[-] \textit{Which similarity function is the most adapted for boundary retrieval in our context?}
\item[-] \textit{Which weighting kernel is the most adapted for boundary retrieval in our context?}
\item[-] \textit{Which penalty function is the most adapted for boundary retrieval in our context?}
\end{itemize}

Each question is addressed sequentially, and the conclusion of each question serves as the basis to study the next ones.

\subsubsection{Train/test datasets}
\label{sec:train_test_datasets}
These questions are addressed by comparing several parameters in a train/test fashion: a subset of the SALAMI dataset, called ``SALAMI-train'', is used to evaluate several parameters, and the best one in this subset is evaluated on the remainder of the SALAMI dataset, called ``SALAMI-test'', and on the entire RWC Pop dataset. The division between SALAMI-train and SALAMI-test is defined by~\cite{ullrich2014boundary}, based on the MIREX evaluation dataset. The details are available online\endnote{\href{https://jan-schlueter.de/pubs/2014_ismir/}{jan-schlueter.de/pubs/2014\_ismir/}}, and are uploaded along with experimental Notebooks on the open-source dataset\endnote{\label{noteCode}\href{https://gitlab.imt-atlantique.fr/a23marmo/autosimilarity_segmentation/-/tree/TISMIR}{https://gitlab.imt-atlantique.fr/a23marmo/\\autosimilarity\_segmentation/-/tree/TISMIR}}. The SALAMI-train dataset contains 849 songs, and the SALAMI-test dataset contains 485 songs\endnote{There is a slight difference in the number of songs in the dataset due to some songs missing in our version of the SALAMI dataset.}. The entire RWC Pop dataset contains 100 songs, resulting in a total of 585 songs for testing.

\subsubsection{Self-similarity Matrices} 
\begin{table*}[htb!]
  \centering
  \begin{tabular}{llllllll}
  \hline
  Self-similarity & $\Pz$ & $\Rz$ & $\Fz$ & $\Po$ & $\Ro$ & $\Fo$ \\ \hline
  Cosine &  \textbf{50.83\%} & 30.82\% & 36.77\% & 62.80\% & 37.72\% & 45.19\% \\
  Autocorrelation & 32.59\% &  \textbf{64.69\%} & 41.30\% & 42.10\% &  \textbf{83.73\%} & 53.41\% \\
  RBF & 50.27\% & 45.38\% &  \textbf{45.84\%} &  \textbf{64.79\%} & 58.81\% & \textbf{59.30\%} \\ \hline
  \end{tabular}
  \caption{Boundary retrieval performance with the different self-similarities on the train dataset. (Full kernel, no penalty function.)}
  \label{table:BTF_full_nopen_train}
\end{table*}

\begin{table*}[htb!]
  \centering
  \begin{tabular}{llllllll}
  \hline
  Dataset & $\Pz$ & $\Rz$ & $\Fz$ & $\Po$ & $\Ro$ & $\Fo$ \\ \hline
  SALAMI - test & 48.52\% & 48.65\% & 46.68\% & 62.76\% & 63.09\% & 60.51\% \\
  RWC Pop & 60.72\% & 53.61\% & 56.01\% & 77.68\% & 67.62\% & 71.09\% \\ \hline
  \end{tabular}
  \caption{Boundary retrieval performance with the RBF self-similarity, on both test datasets. (Full kernel, no penalty function.)}
  \label{table:BTF_full_nopen_test}
\end{table*}

\begin{figure}[htb!]
\centering
\begin{subfigure}{0.95\columnwidth}
  \includegraphics[width=\columnwidth]{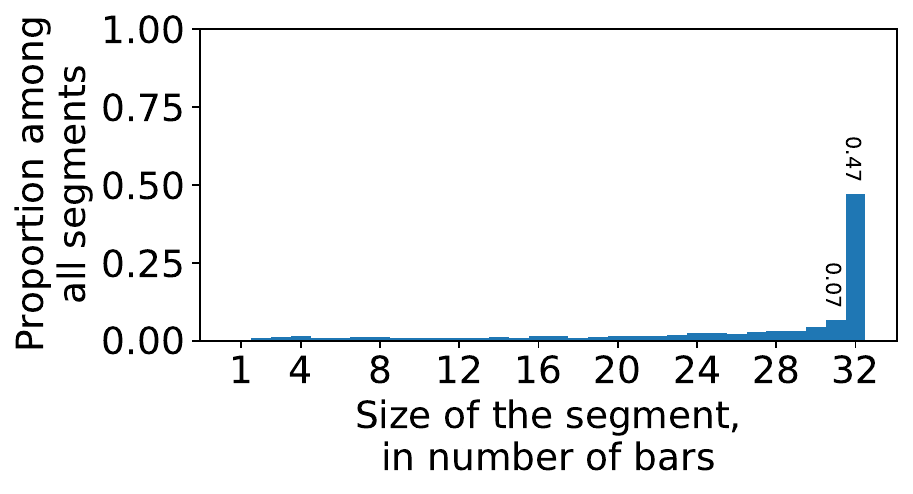}
  \caption{Cosine self-similarity.}
\end{subfigure}
\quad
\begin{subfigure}{0.95\columnwidth}
      \vspace{10pt}
  \includegraphics[width=\columnwidth]{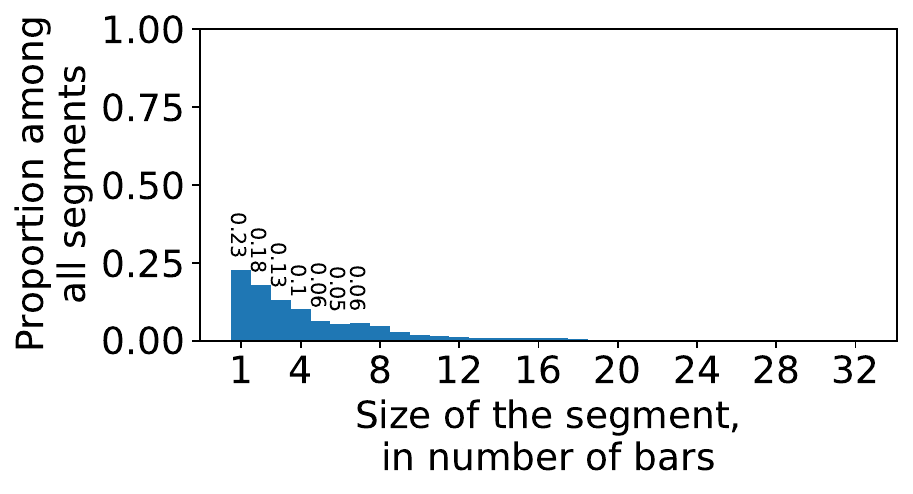}
  \caption{Autocorrelation self-similarity.}
  \end{subfigure}
  \quad
  \begin{subfigure}{0.95\columnwidth}
      \vspace{10pt}
    \includegraphics[width=\columnwidth]{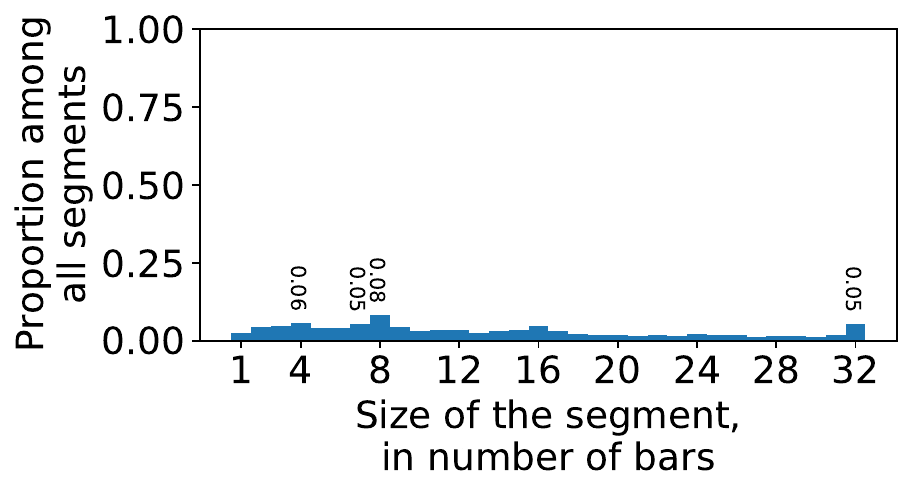}
    \caption{RBF self-similarity.}
    \end{subfigure}
    \vspace{0.1cm}
\caption{Distribution of the segment sizes, with the full kernel, according to the self-similarity matrix. Results on the SALAMI-train dataset.}
\label{fig:BTF_full_distrib_nopen_rwc}
\end{figure}

Firstly, we study the impact of the design of the similarity function 
on the performance of the CBM algorithm. To do so, we use the CBM algorithm with the full kernel, as it does not need the fitting of the number of bands, and 
we do not use a penalty function. The boundary retrieval performance is presented in the Table~\ref{table:BTF_full_nopen_train} for the train dataset.

The RBF self-similarity is the best-performing self-similarity in terms of F-measure (with both tolerances), hence suggesting a better boundary estimation in average than the other similarity functions. The results obtained with the RBF similarity function on the test datasets are presented in Table~\ref{table:BTF_full_nopen_test}. 

The precision/recall trade-offs depend on the self-similarity matrices, and deserve to be studied to give further information on the quality of the estimated segmentations.~The Cosine self-similarity exhibits a higher precision than recall on average, which suggests an under-segmentation, \ie estimating too few boundaries.~
Conversely, the Autocorrelation self-similarity results in a higher recall than precision, suggesting over-segmentation. The RBF self-similarity performance is more balanced between both metrics.

These conclusions can be confirmed by studying the distribution of the sizes of the estimated segments, as presented in Figure~\ref{fig:BTF_full_distrib_nopen_rwc} on the SALAMI-train dataset. These distributions must be compared with the distribution of segment sizes in the annotation, presented in Figure~\ref{fig:distrib_barwise_segment_sizes_cbm_annotation}.

The distribution of segment sizes with the RBF self-similarity is visually the closest one to the distribution in annotation, which we confirm numerically by studying the Kullback-Leibler (KL) divergences between the distribution of the sizes of the estimated segments and of the annotated ones. The KL-divergences are respectively equal to 2.25, 0.85 and 0.35 for the Cosine, Autocorrelation and RBF similarity functions. Again, this suggests that the RBF similarity function is the most adapted.


\subsubsection{Block Weighting Kernels}
\begin{figure*}[htb!]
\centering
\begin{subfigure}{0.95\columnwidth}
  \includegraphics[width=\columnwidth]{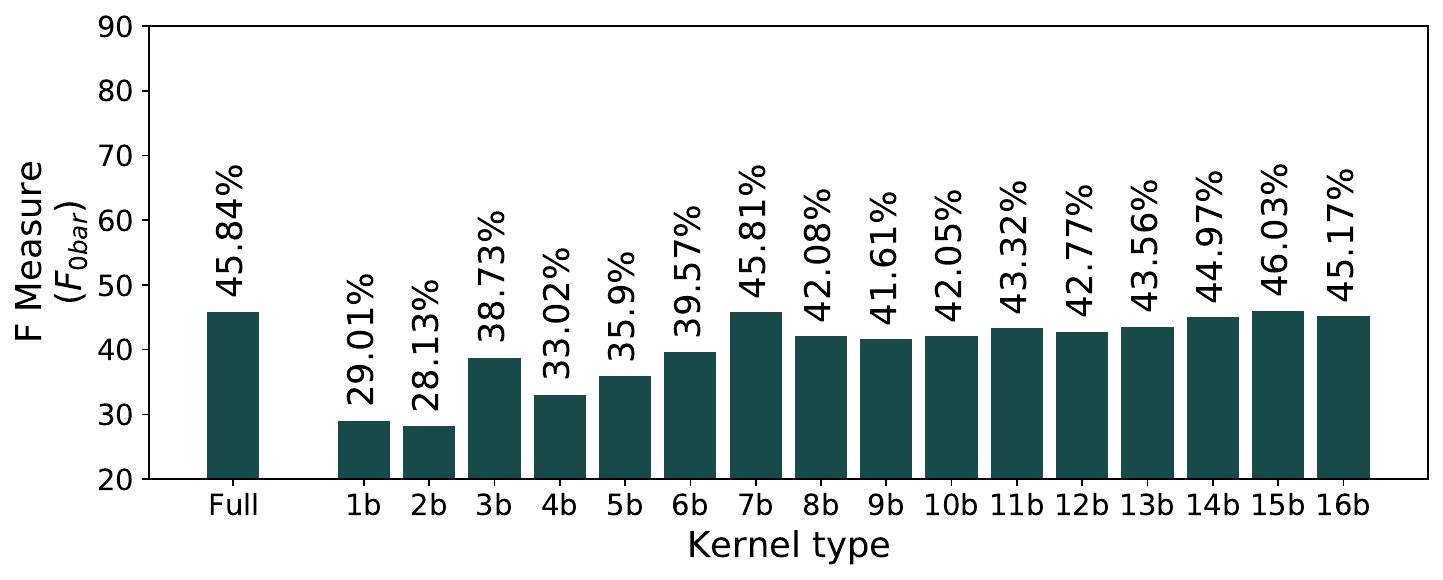}
  \caption{$\Fz$}
\end{subfigure}
\quad
\begin{subfigure}{0.95\columnwidth}
  \includegraphics[width=\columnwidth]{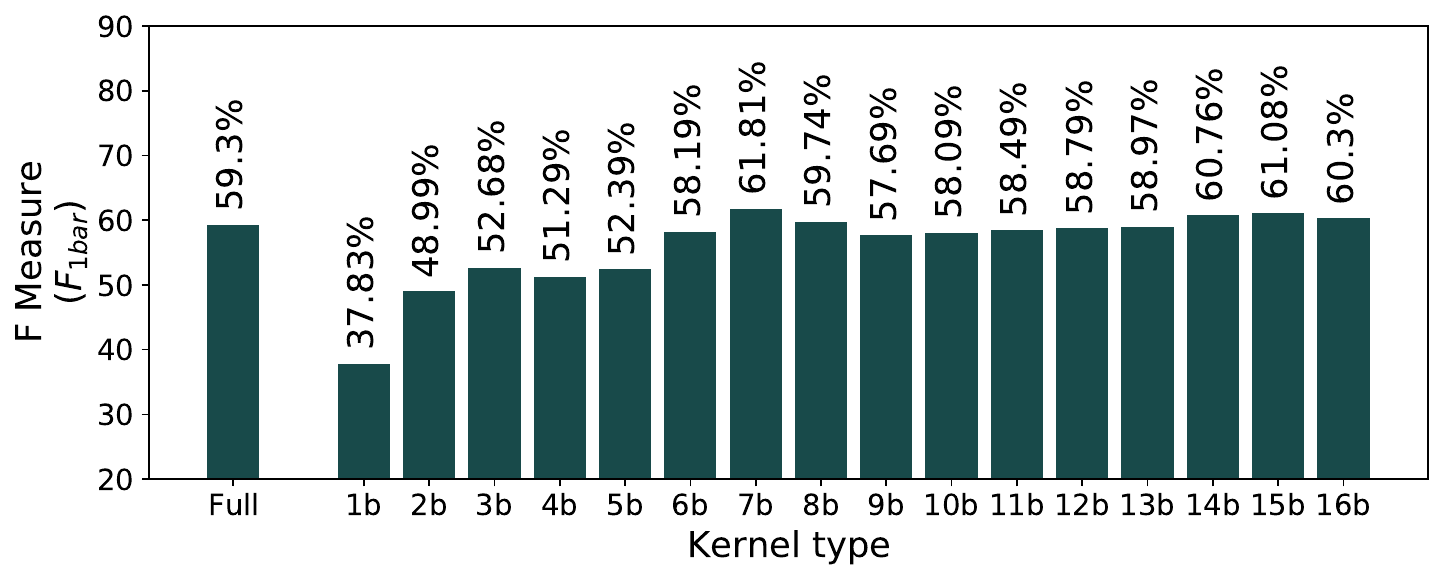}
  \caption{$\Fo$}
  \end{subfigure}
  \vspace{5pt}
\caption{Boundary retrieval performance (F-measures only) according to the full and band kernels (with different number of bands). Results on the train dataset with RBF self-similarity matrices.}
\vspace{5pt}
\label{fig:BTF_bands_nopen_train}
\end{figure*}

Secondly, an important parameter in the CBM algorithm is the design of the kernel.~We thus compare the full kernel with band kernels, the number of bands varying from 1 to 16 bands. Results on the SALAMI-train dataset, computed on the RBF self-similarity matrices, and focusing on the F-measures, are presented in Figure~\ref{fig:BTF_bands_nopen_train}. The 7-band kernel stands out as the best-performing kernel, even if performance is close to the 15-band kernel.

\begin{figure*}[htb!]
\centering
\begin{subfigure}{0.95\columnwidth}
  \includegraphics[width=\columnwidth]{figs/segments_sizes/segments_distrib_train_full_nopen_rbf.pdf}
  \caption{Full kernel.}
  \vspace{3mm}
\end{subfigure}
\quad
\begin{subfigure}{0.95\columnwidth}
  \includegraphics[width=\columnwidth]{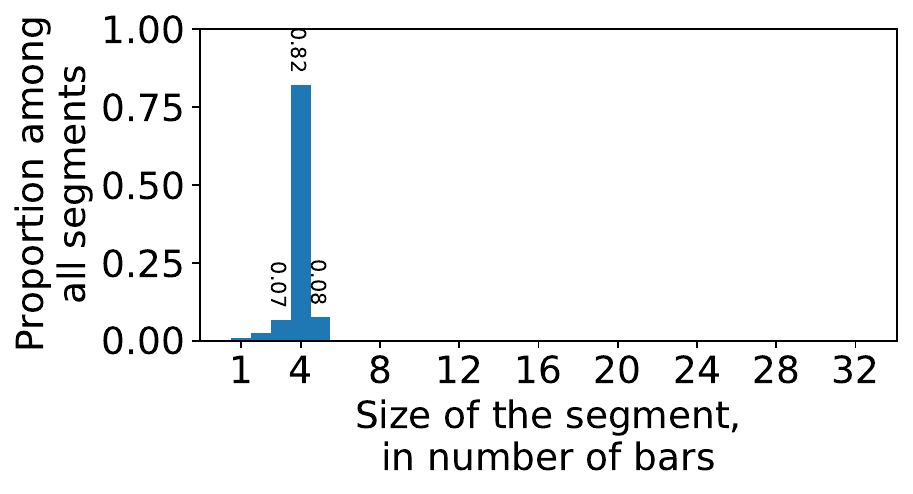}
  \caption{3-band kernel.}  
  \vspace{3mm}
  \end{subfigure}
  \quad
  \begin{subfigure}{0.95\columnwidth}
    \includegraphics[width=\columnwidth]{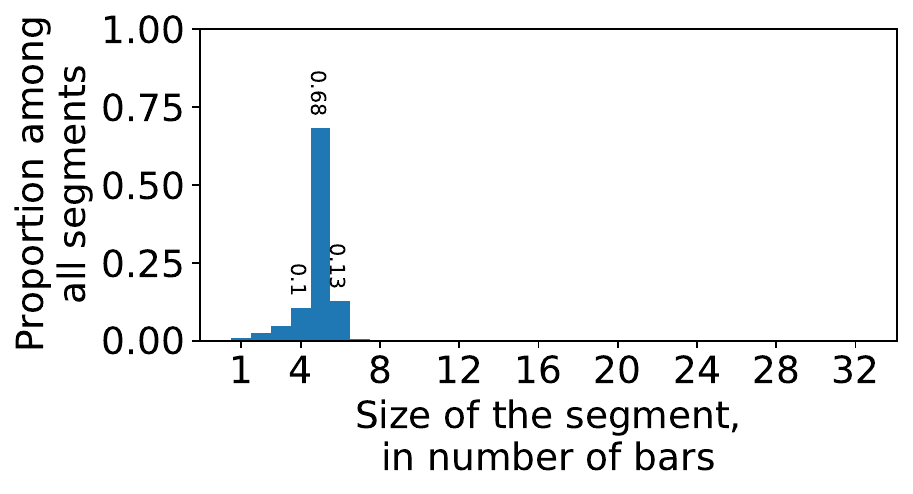}
    \caption{4-band kernel.}
      \vspace{3mm}
    \end{subfigure}
    \quad
    \begin{subfigure}{0.95\columnwidth}
      \includegraphics[width=\columnwidth]{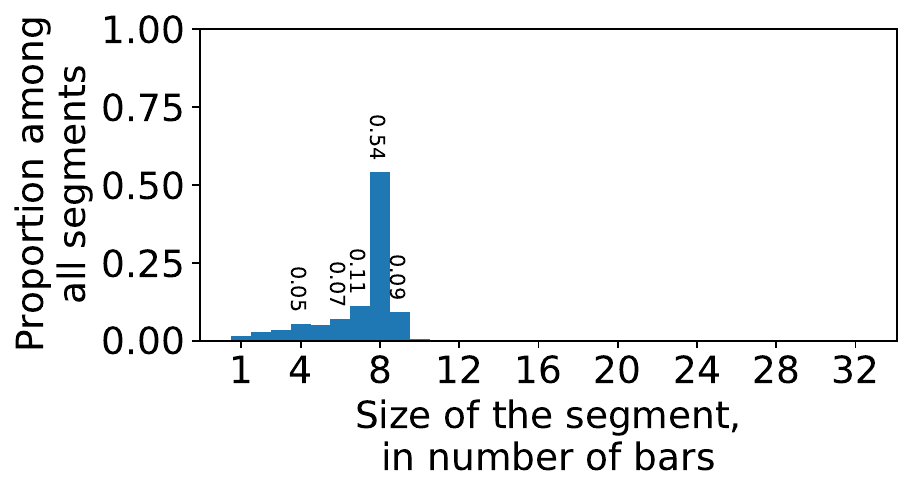}
      \caption{7-band kernel.}
        \vspace{3mm}
    \end{subfigure}
    \quad
    \begin{subfigure}{0.95\columnwidth}
      \includegraphics[width=\columnwidth]{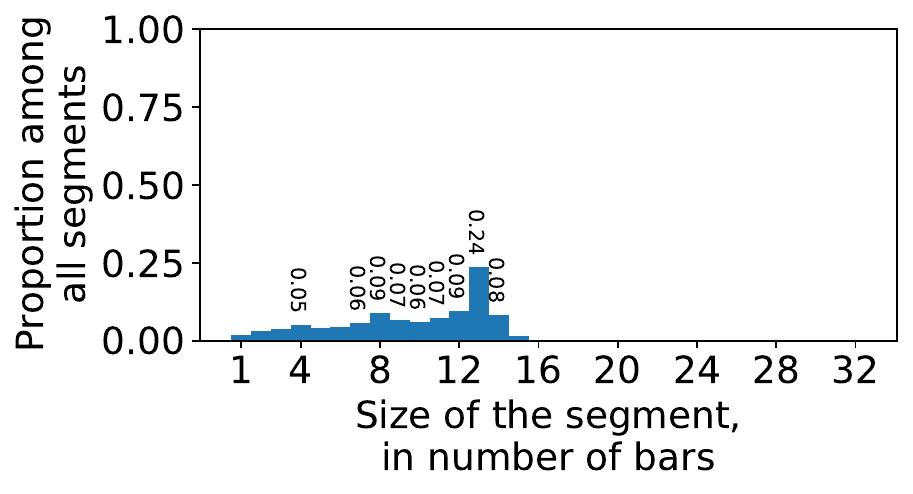}
      \caption{12-band kernel.}
        \vspace{2mm}
      \end{subfigure}
      \quad
      \begin{subfigure}{0.95\columnwidth}
        \includegraphics[width=\columnwidth]{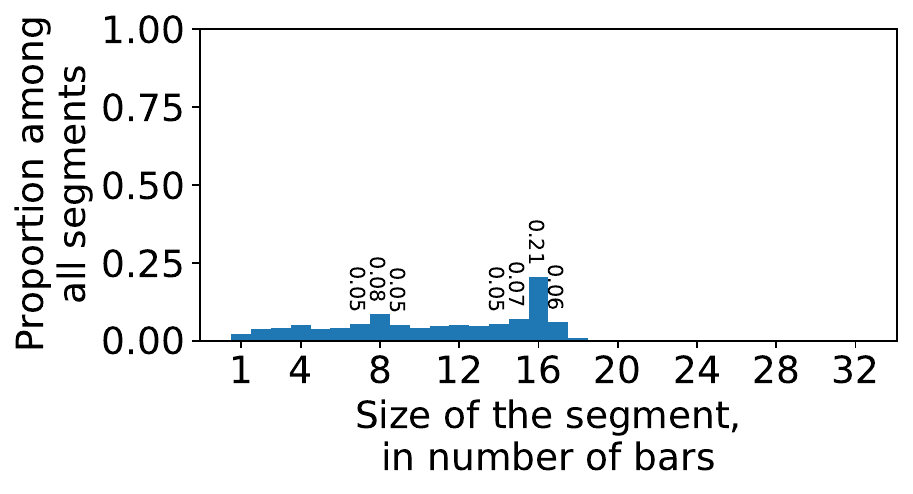}
        \caption{15-band kernel.}
          \vspace{2mm}
        \end{subfigure}
\caption{Distribution of estimated segment sizes, according to different kernels, on the train dataset.}
\vspace{10pt}
\label{fig:BTF_kernel_distrib_nopen_train}
\end{figure*}

\begin{table*}[htb!]
  \centering
  \begin{tabular}{llllllll}
  \hline
  Dataset & $\Pz$ & $\Rz$ & $\Fz$ & $\Po$ & $\Ro$ & $\Fo$ \\ \hline
  SALAMI - test & 37.24\% & 59.80\% & 44.33\% & 50.38\% & 80.52\% & 59.88\% \\
  RWC Pop & 59.41\% & 68.19\% & 62.82\% & 75.53\% & 86.56\% & 79.81\% \\ \hline
  \end{tabular}
  \caption{Boundary retrieval performance with the 7-band kernel, on both test datasets. (RBF self-similarity, no penalty function.)}
  \label{table:BTF_7b_nopen_test}
\end{table*}

The differences in performance between the different kernels may be explained by the Figure~\ref{fig:BTF_kernel_distrib_nopen_train}, which presents the distribution of segment sizes according to the number of bands. The 7-band kernel leads to a majority of estimated segments of size 8 (more than 50\%, twice as much as in the annotation), which is the most common segment size in the annotation, while the 15-band kernel mostly computes segments of size 16, and the full kernel is well distributed across the different segment sizes. The annotations are mostly composed of segments of size 8, then 4, 12 and 16. Hence, while the 7-band kernel does not accurately represent the annotation, it obtains better boundary retrieval performance than the other ones, indicating that this latter distribution is beneficial to the boundary estimation overall.

As an additional conclusion, the number of bands in the kernel largely influences the distribution of segment sizes, in particular the most frequent segment size. As a general trend, it seems that a kernel with $v$ bands favors segments of size $v + 1$. We assume that this behavior stems from the fact that, for a $v$-band kernel and a large segment of size $n > v$, the number of elements equal to 0 is large, but the normalization remains adapted to kernels with $n^2$ values. We found in practice that this effect could be dampened by normalizing the score associated with each kernel by the number of nonzero values plus the number of elements in the diagonal instead of the size of the kernel, as in~\citep{shiu2006similarity}, but this resulted in all kernels performing similarly than the full kernel, hence less performing than the 7-band one.

Finally, as the 7-band kernel is the best-performing one, we fixed this kernel for both test datasets. Results obtained with this kernel are presented in Table~\ref{table:BTF_7b_nopen_test}.

\subsubsection{Penalty Functions} 
\begin{table*}[htb!]
\centering
\begin{tabular}{lllllllll}
\hline
\multicolumn{2}{l} {Penalty function} & Best $\lambda$ & $\Pz$ & $\Rz$ & $\Fz$ & $\Po$ & $\Ro$ & $\Fo$ \\ \hline
\multicolumn{2}{l}{Without penalty}    & - & 40.26\% & 57.38\% & 45.81\% & 54.26\% & \textbf{77.67\%} & 61.81\%  \\ \hline
\multirow{3}{*}{Target deviation} & $\alpha = \frac{1}{2}$ & 0.01  & 40.38\% & 57.36\% & 45.88\% & 54.37\% & 77.57\% & 61.84\% \\ \cline{2-9}
 & $\alpha = 1$ & 0.01 & 40.45\% & 56.98\% & 45.81\% & 54.61\% & 77.20\% & \textbf{61.89\%} \\ \cline{2-9}
  & $\alpha = 2$ & 0.01 & 39.75\% & 54.32\% & 44.43\% & \textbf{54.93\%} & 75.31\% & 61.46\% \\ \hline
  \multicolumn{2}{l}{Modulo 8} & 0.04 & \textbf{41.04\%} & \textbf{58.34\%} & \textbf{46.63\%} & 54.25\% & 77.44\% & 61.72\% \\ \hline
\end{tabular}
\caption{Boundary retrieval performance depending on the penalty function, for the SALAMI-train dataset, with the RBF self-similarity and the 7-band kernel.}
\label{table:penalty_cbm_train}
\vspace{8pt}

\end{table*}

\begin{table*}[htb!]
  \centering
  \begin{tabular}{llllllll}
  \hline
  Dataset & $\Pz$ & $\Rz$ & $\Fz$ & $\Po$ & $\Ro$ & $\Fo$ \\ \hline
  SALAMI - test & 38.36\% & 60.96\% & 45.44\% & 50.76\% & 80.51\% & 60.09\% \\
  RWC Pop & 62.11\% & 70.05\% & 65.17\% & 77.35\% & 86.95\% & 81.02\% \\ \hline
  \end{tabular}
  \caption{Boundary retrieval performance with the modulo 8 penalty function ($\lambda = 0.04$), on both test datasets (RBF self-similarity, 7-band kernel).}
  \label{table:penalty_cbm_test}
\end{table*}

Finally, the last experiments focus on the penalty functions. In this set of experiments, we compare the target deviation functions, with $\alpha \in \{0.5, 1, 2\}$, with the modulo 8 function. The CBM algorithm is parametrized with the 7-band kernel, and is applied on the RBF self-similarity matrices. The parameter $\lambda$, balancing the penalty function, takes values between $\frac{1}{100}$ and $\frac{2}{10}$, with a step of $\frac{1}{100}$. This parameter is fitted on the SALAMI-train dataset. Results are presented in Table~\ref{table:penalty_cbm_train}.

The modulo 8 function appears to slightly improve boundary retrieval performance for the metrics with a tolerance of 0 bar, indicating a more accurate estimation, but results with a tolerance of 1 bar are not strongly impacted by the choice of the penalty function. Results are close between the different penalty functions, except for the target deviation with a large $\alpha$, which results in worse performance than the other conditions.

Overall, it seems that the modulo 8 function is the most adapted penalty function to estimate segments more accurately. Hence, we use this penalty function for the test results, presented in Table~\ref{table:penalty_cbm_test}. Parameter $\lambda = 0.04$, as optimized on the SALAMI-train dataset.

\subsubsection{Experimental Conclusions}
In light of these results, we finally sum up the situation regarding the choice of settings in the CBM algorithm.

\begin{enumerate}
\item \textit{In our context, the RBF self-similarity matrix is the most adapted self-similarity matrix for boundary retrieval.}

\item \textit{In our context, the 7-band kernel is the most adapted kernel for boundary retrieval.}

\item \textit{In our context, the modulo 8 penalty function is the most adapted penalty function for boundary retrieval.}
\end{enumerate}

\subsubsection{Metrics With Tolerance in Absolute Time}
As mentioned in Section~\ref{sec:metrics}, standard metrics for boundary retrieval performance consider the tolerance in absolute time (\eg $\Fzf$ and $\Fth$ metrics), while we opted for boundary-aligned metrics in our experiments. Hence, Table~\ref{table:best_results_tol_time_bar} compares the boundary retrieval performance obtained when the tolerance is defined relatively to the bars and in absolute time, which allows to compare with State-of-the-Art algorithms. Boundary retrieval performance is almost equivalent on the RWC Pop dataset, and slightly altered for the metrics with short tolerances on the SALAMI dataset ($\Fz$ and $\Fzf$). These discrepancies may be explained by the less precise downbeat alignment of annotations in the SALAMI dataset, presented in Table~\ref{table:ref_aligned_downbeats}. Overall though, results remain similar, which tends to confirm the hypothesis of~\cite{ong2005semantic} that a tolerance of 3 seconds corresponds approximately to a tolerance of 1 bar.

\begin{table}[tbh!]
\centering
\begin{tabular}{llllll}
\hline
Dataset & $\Fz$ & $\Fzf$ & $\Fo$ & $\Fth$ \\ \hline
SALAMI-test &  45.44\% & 42.00\% & 60.09\% & 60.61\% \\ \hline
RWC Pop & 65.17\% & 64.44\% & 81.02\% & 80.64\%  \\ \hline
\end{tabular}
\caption{Boundary retrieval performance, comparing the F-measures with tolerance expressed barwise and in absolute time.}
\label{table:best_results_tol_time_bar}
\end{table}

\subsection{Comparison with State-of-the-Art Algorithms}
We compare the boundary retrieval performance obtained by the Foote-TF (introduced in Section~\ref{sec:foote_details}) and the CBM algorithms with State-of-the-Art algorithms. The performance of the CBM algorithm is obtained using the hyperparameters learned in Section~\ref{sec:parametrization}.

This work considers seven different algorithms as State-of-the-Art, categorized as either unsupervised or supervised algorithms, \ie algorithms that either estimate boundaries without the use of training examples or analyze annotated examples before making predictions: four unsupervised algorithms~\citep{foote2000automatic, mcfee2014analyzing, serra2014unsupervised, mccallum2019unsupervised} and three supervised algorithms~\citep{grill2015cnn, wang2021supervised, salamon2021deep}. We additionally use previous work on the CBM algorithm~\citep{marmoret2022barwise} as baseline.

All State-of-the-Art algorithms use beat-aligned features, except~\cite{grill2015cnn} which uses a fixed hop length and~\cite{wang2021supervised} which uses downbeat-aligned features. In details, results for~\cite{foote2000automatic, mcfee2014analyzing, serra2014unsupervised} are computed with the \textit{MSAF} toolbox~\citep{msaf}, and realigned on downbeats in post-processing. Results for the CNN~\citep{grill2015cnn} are extracted from the 2015 MIREX contest. Results for~\cite{mccallum2019unsupervised, wang2021supervised, salamon2021deep} are copied from the respective articles.

\begin{figure*}[tb!]
\centering
\begin{subfigure}{0.95\columnwidth}
  \includegraphics[width=\columnwidth]{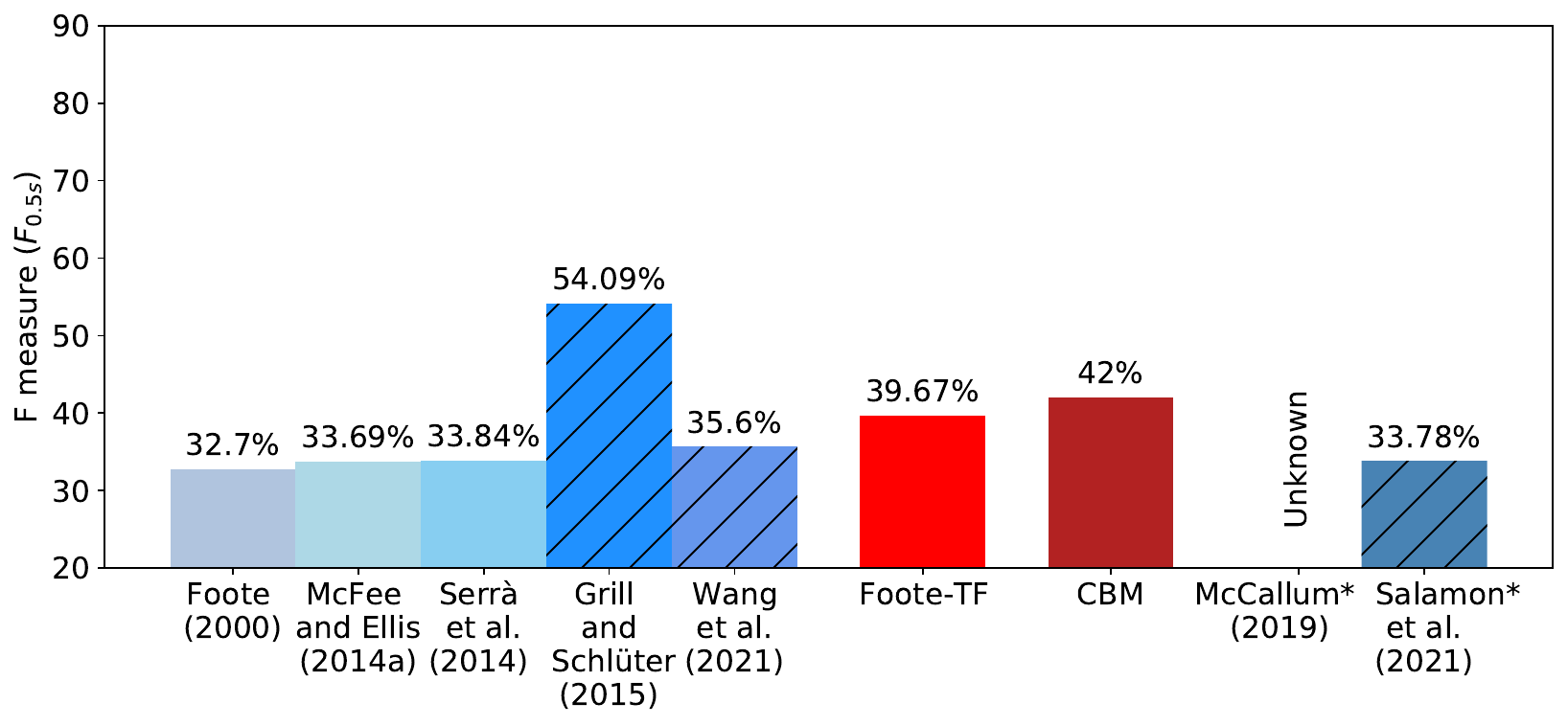}
  \caption{$\Fzf$}
\end{subfigure}
\begin{subfigure}{0.95\columnwidth}
  \includegraphics[width=\columnwidth]{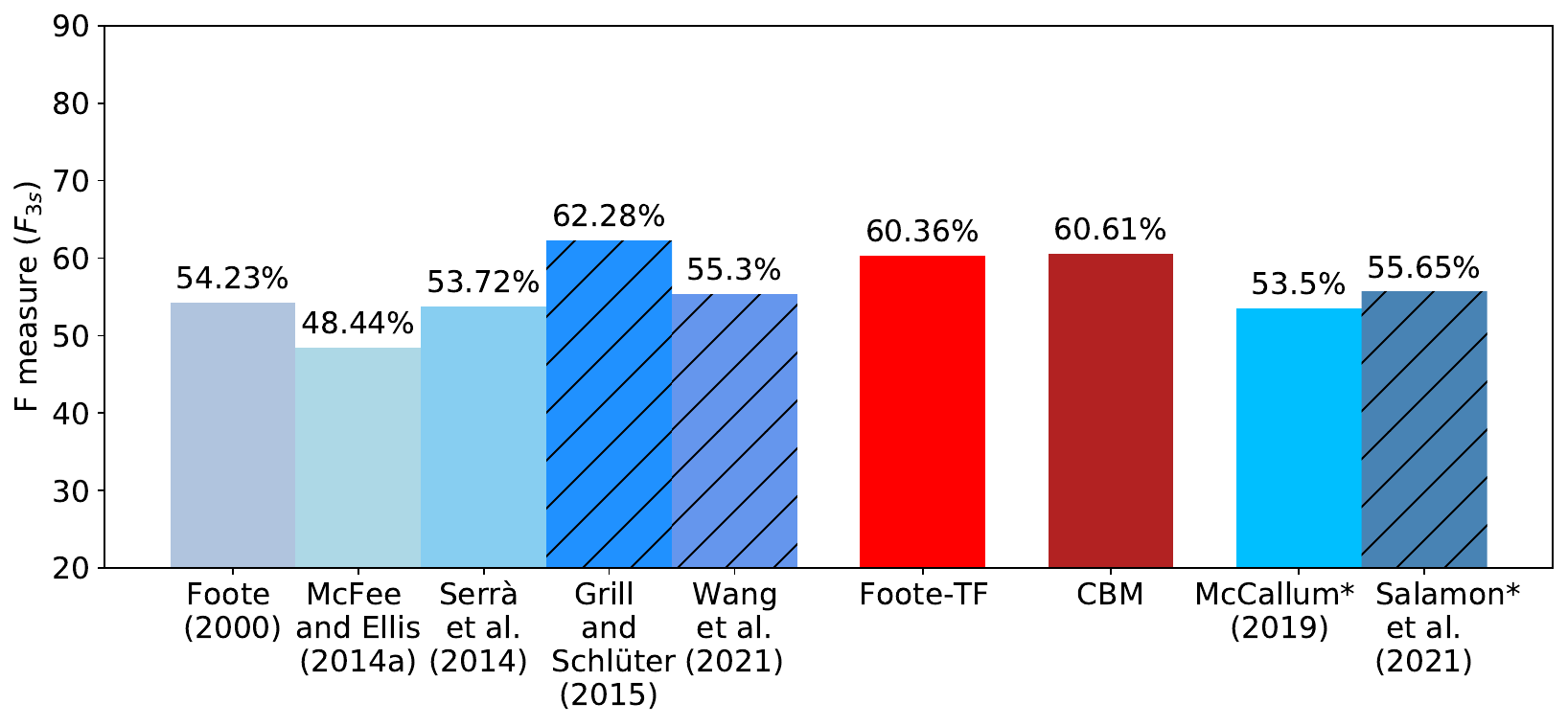}
  \caption{$\Fth$}
  \end{subfigure}
  \caption{Boundary retrieval performance of the CBM algorithm on the SALAMI dataset, compared to State-of-the-Art algorithms. Hatched bars correspond to supervised algorithms. The star * represents algorithms were the evaluation subset is not exactly the same as ours, thus preventing accurate comparison.}
\label{fig:comparison_SOTA_salami}
\end{figure*}
\begin{figure*}[htb!]
\centering
\begin{subfigure}{0.95\columnwidth}
  \includegraphics[width=\columnwidth]{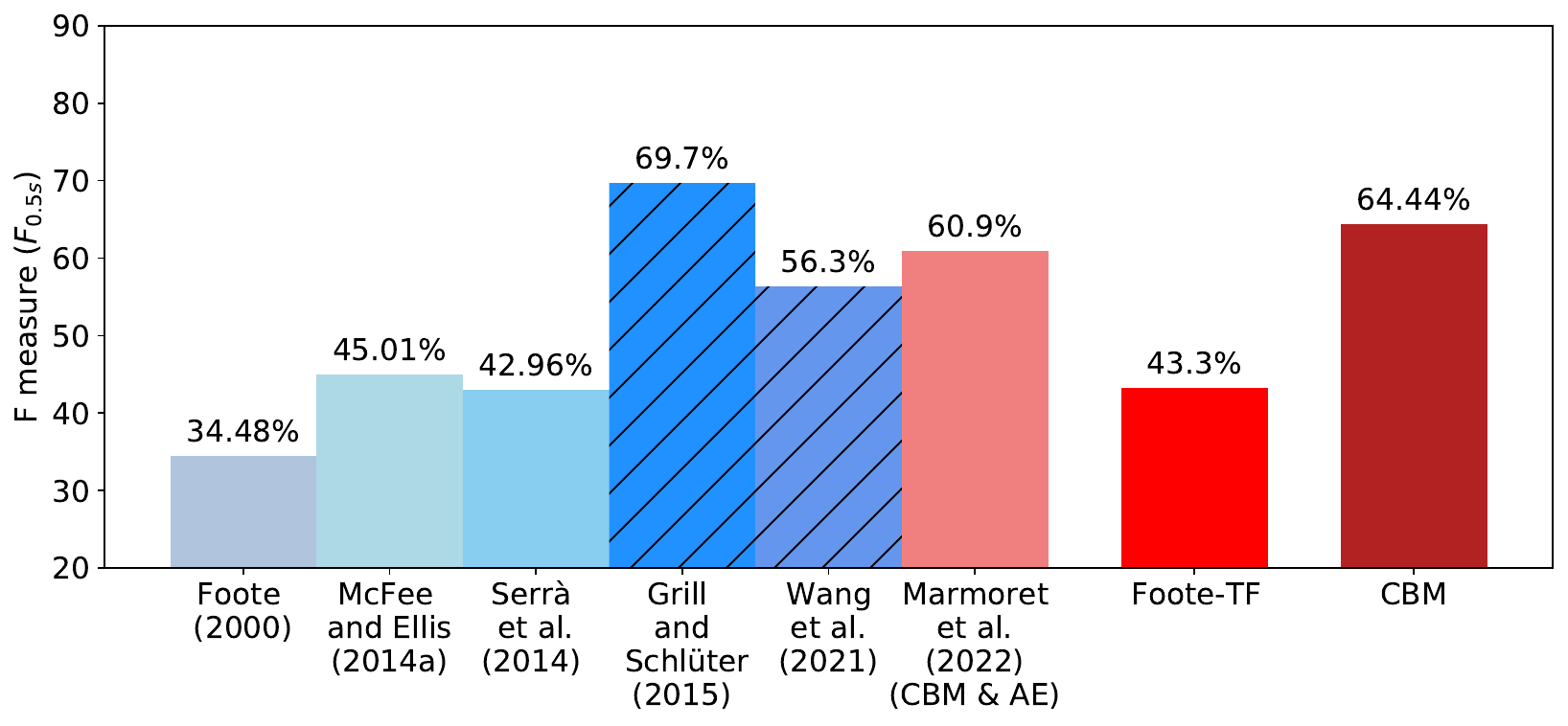}
  \caption{$\Fzf$}
\end{subfigure}
\begin{subfigure}{0.95\columnwidth}
  \includegraphics[width=\columnwidth]{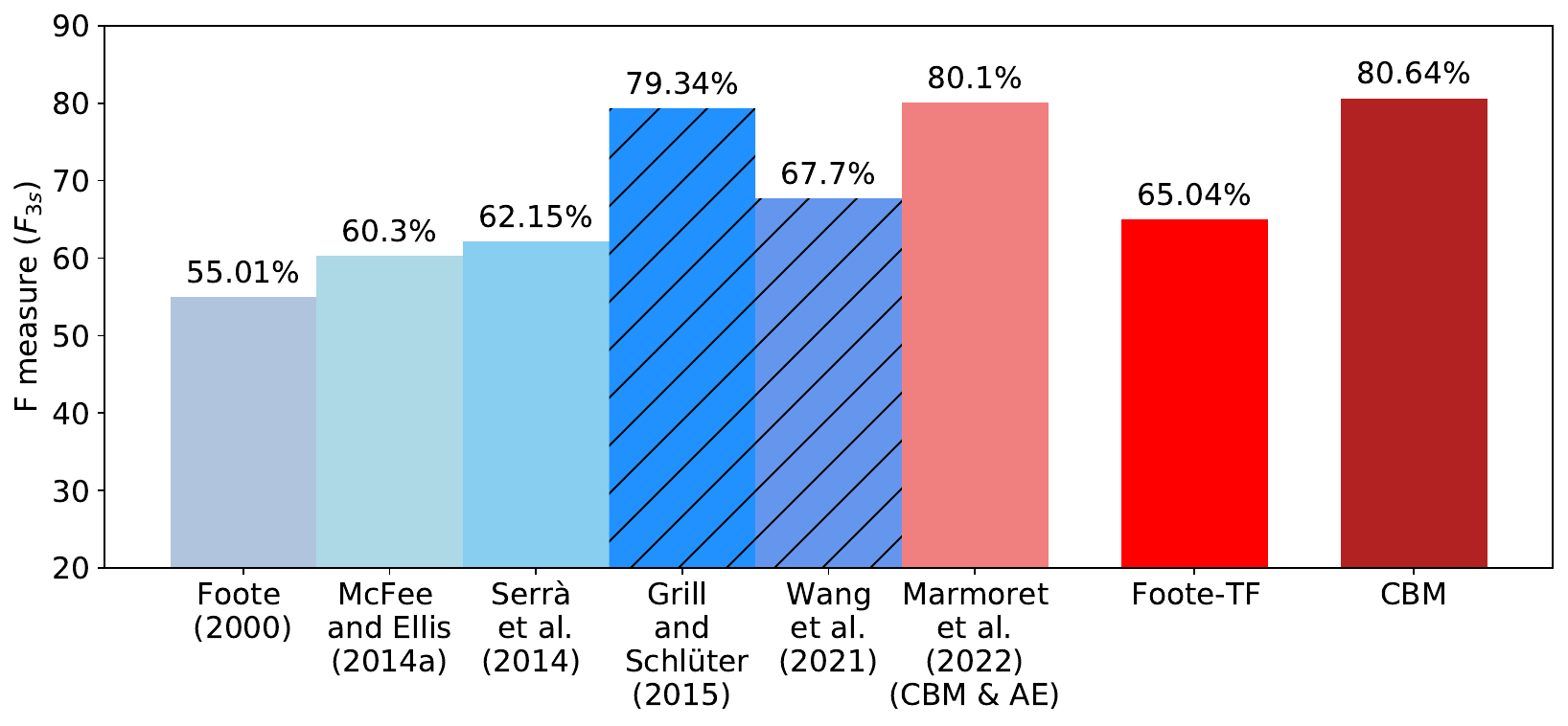}
  \caption{$\Fth$}
  \end{subfigure}
  \caption{Boundary retrieval performance of the CBM algorithm on the RWC Pop dataset, compared to State-of-the-Art algorithms. Hatched bars correspond to supervised algorithms.}
\label{fig:comparison_SOTA_rwc}
\end{figure*}

Figures~\ref{fig:comparison_SOTA_salami} and~\ref{fig:comparison_SOTA_rwc} compare the results obtained with the CBM and the Foote-TF algorithms with those of the State-of-the-Art algorithms\endnote{A careful reader will notice that results of~\citep{foote2000automatic} in Figures~\ref{fig:comparison_SOTA_salami} and~\ref{fig:comparison_SOTA_rwc} are not exactly the sames than those obtained in Tables~\ref{table:foote_beat_bar_salami} and~\ref{table:foote_beat_bar_rwc}.~We explain these discrepancies by the use of different beatwise estimation algorithms: the \textit{madmom} toolbox for Figures~\ref{fig:comparison_SOTA_salami} and~\ref{fig:comparison_SOTA_rwc}, and the original implementation in the \textit{MSAF} toolbox for Tables~\ref{table:foote_beat_bar_salami} and~\ref{table:foote_beat_bar_rwc}.}. In this comparison, the CBM algorithm globally outperforms the other unsupervised segmentation methods, most of the supervised algorithms, and is competitive for the metric $\Fth$ with the global (supervised) State-of-the-Art~\citep{grill2015cnn}. These results are promising and show the potential of the CBM algorithm, which is performing well despite its relative simplicity. Additionally, results of the CBM algorithm are on par with those of Foote-TF on the SALAMI dataset, showcasing interest for the Barwise TF representation.\endnote{We recall here that the Barwise TF matrix of the Foote-TF algorithm uses the parameters of~\citep{foote2000automatic}, \ie Chromagrams and cosine autosimilarity matrix. Still, the Foote's algorithm implementation in~\citep{msaf} also uses pre-processing steps which benefits the algorithm, and could benefit the CBM algorithm too. Studying these types of bridges between both algorithms is left to future work.}

\subsection{CBM: Barscale \textit{vs.} Beatscale}

\begin{table*}[htb!]
\centering
\begin{tabular}{lllllll}
\hline
SALAMI & $\Pzf$ & $\Rzf$ & $\Fzf$ & $\Pth$ & $\Rth$ & $\Fth$ \\ \hline
Beatwise (cosine, 63-band kernel) & \textbf{35.90\%} & 41.61\% & 37.36\% & \textbf{55.75\%} & 64.52\% & 58.03\% \\
Barwise (RBF, 7-band kernel) & 34.49\% & \textbf{54.56\%} & \textbf{41.04\%} & 50.70\% & \textbf{80.78\%} & \textbf{60.51\%} \\ \hline
\end{tabular}
\caption{CBM algorithm, performed on Barwise TF matrix \textit{vs.} Beatwise TF matrix, on the SALAMI-test dataset. For fairer comparison, results at both scales are computed without penalty function.}
\label{table:barwise_beatwise_salami}
\end{table*}

\begin{table*}[htb!]
\centering
\begin{tabular}{lllllll}
\hline
RWC Pop & $\Pzf$ & $\Rzf$ & $\Fzf$ & $\Pth$ & $\Rth$ & $\Fth$ \\ \hline
Beatwise (cosine, 63-band kernel) & 46.22\% & 44.38\% & 44.57\% & 72.54\% & 68.85\% & 69.51\% \\ 
Barwise (RBF, 7-band kernel) & \textbf{59.09\%} & \textbf{67.13\%} & \textbf{62.28\%} & \textbf{75.17\%} & \textbf{85.90\%} & \textbf{79.47\%} \\ \hline
\end{tabular}
\caption{CBM algorithm, performed on Barwise TF matrix \textit{vs.} Beatwise TF matrix, on RWC Pop. For fairer comparison, results at both scales are computed without penalty function.}
\label{table:barwise_beatwise_rwc}
\end{table*}

In order to distinguish the impact of barwise-alignment and of the CBM algorithm itself on the segmentation results, we compare results obtained on the Barwise TF matrix, presented above, with results obtained on the ``Beatwise TF matrix'', \ie the equivalent of the Barwise TF matrix at the beat scale. Following the intuition that most bars in Western modern music are composed of 4 beats per bar, the Beatwise TF matrix is sampled with $T = 24$ samples per beat, \ie $\frac{96}{4}$. Beats are estimated with the algorithm of~\cite{bock2019multi}, implemented in the \textit{madmom} toolbox, as in Section~\ref{sec:experiments_barwise_sota}. For fairer comparison, the results at both beatwise and barwise scales are computed without penalty function. Corresponding results are presented in Tables~\ref{table:barwise_beatwise_salami} and~\ref{table:barwise_beatwise_rwc}.

Results confirm that barwise-alignment is beneficial for the performance of the CBM algorithm, especially on the RWC Pop dataset, but the CBM algorithm with beat-aligned features obtains similar or better performance than the unsupervised State-of-the-Art algorithms~\citep{foote2000automatic, mcfee2014analyzing, serra2014unsupervised, mccallum2019unsupervised}, and is competitive with the supervised algorithm~\citep{wang2021supervised} on both datasets for the metric $\Fth$.

\section{Conclusions}
In this article, the CBM algorithm has been thouroughly presented to perform Music Structure Analysis on audio signals, where boundaries between musical sections are computed by maximizing the homogeneity of each segment composing the segmentation, using dynamic programming under a penalty function. Moreover, a barwise processing of music is shown to increase segmentation performance, using the Barwise TF matrix. This work has also investigated several metrics to represent similarities between pairs of bars in a song.~
While the CBM algorithm has room for improvement, it achieves a level of performance which is competitive to the State-of-the-Art, and therefore appears as a meaningful approach to investigate a variety of music representations without needing large collections of training data. 

The design of the kernel clearly impacts the boundary retrieval performance. Hence, future work could focus on studying alternative types of kernels.~The kernel values could depend on the particular song or dataset considered, or follow particular statistical distribution.~Of particular interest could be the learning of such kernels instead of an (empirical) definition.~These latter comments are also valid for the penalty functions, whose values were set quite empirically, and which would benefit from deeper investigation.~The number of bands in the weighting kernels seems to enforce particular segment sizes.~This effect can be mitigated with normalization, or, conversely, further exploited, for instance by using different kernels concurrently, each one accounting for a different level of structure, hence studying segmentation hierarchically.

Weighting kernels presented in this article focus on the homogeneity of each segment, but other kernels could be considered in order to account for repetition in the song. In fact, the proposed framework is highly customizable with respect to weighting kernels, and could be adapted to the expected shape of segments. In particular, we expect that bridging this work with previous work~\citep{foote2000automatic} could further enhance performance.

\section{Reproducibility}
All the code used in this article is contained in the open-source toolbox~\citep{marmoret2022as_seg}, along with experimental Notebooks used to compute the experimental results\textsuperscript{\ref{noteCode}}.

\IfFileExists{\jobname.ent}{
   \theendnotes
}{
}

\section*{Acknowledgment}
This work is partly supported by ANR JCJC project LoRAiA ANR-20-CE23-0010. 
\bibliography{TISMIRtemplate}

\appendix
\section{Algorithm, in details}
The detailed algorithm, in pseudo-code, is presented hereafter. 
\begin{algorithm}[ht!]
\DontPrintSemicolon
\SetAlgoLined
 \caption{CBM algorithm, computing the optimal segmentation given a score function $\SegScoreNotation()$.}
\KwIn{Bars $\{b_k \in \llbracket 1, B \rrbracket\}$, score function $\SegScoreNotation$}
\KwOut{Optimal segmentation $\BoundSet^* = \{\aBound_i\}$}
\,\\
$\BoundSet^* = \{1,  B + 1\}$\\

$\SetOfAntecedents = [\,]$\\
\tcc{Array storing the optimal antecedents for every bar (empty at initialization).}
\,\\

$\SetOfScores = [0]$\\
\tcc{Array storing the optimal segmentation up to each bar (set to $\SetOfScores[1] = 0$ at initialization).}
\For{$b_k = 2, ..., B+1$}{
    $\aBound = b_k$\\
    \tcc{Considering bar $b_k$ as current boundary.}
    \,\\

      $\aBound_{-1}^* = \underset{1 \leq \aBound_j < \aBound}{\argmax} \, \left(\SetOfScores[\aBound_j] + \SegScoreNotation(\llbracket \aBound_j, \aBound - 1\rrbracket)\right)$\\
      \tcc{Finding the best antecedent for the current boundary $\aBound$ with Equation~\ref{eq:recurrence_property_dynaprog}.}
      \,\\

      $\SetOfAntecedents[\aBound] = \aBound_{-1}^*$\\
      \tcc{Storing the best antecedent for $\aBound$.}
      \,\\

    $\SetOfScores[\aBound] = \SetOfScores[\aBound_{-1}^*] + \SegScoreNotation(\llbracket \aBound_{-1}^*, \aBound - 1 \rrbracket)$\\
    \tcc{Computing and storing the optimal segmentation score up to $\aBound$.}
}

$\aBound = B+1$\\ 
\While{$\SetOfAntecedents[\aBound] \neq 1$}{
\tcc{Recursively backtracking all optimal antecedents, from the last to the first bar.}
\,\\

$\aBound_{-1}^* = \SetOfAntecedents[\aBound]$\\
$\BoundSet^* = \BoundSet^* \cup \{\aBound_{-1}^*\}$\\
\tcc{Searching for the best antecedent and adding it to the optimal segmentation.}
\,\\

$\aBound = \aBound_{-1}^*$\\
\tcc{Iterating the process with the current best antecedent.}
}

\label{alg:dyna_conv_algo}
\end{algorithm}

Practically, the advantage of the algorithm is to be able to store in memory both the optimal antecedent for each bar and the scores of the optimal segmentation up to each bar when they are computed for the first time, respectively denoted as the arrays $\SetOfAntecedents$ and $\SetOfScores$ in Algorithm~\ref{alg:dyna_conv_algo}.

\end{document}